\documentclass{aa}
\usepackage[colorlinks=true,linkcolor=blue,citecolor    = blue]{hyperref}

\usepackage[]{natbib} 
\bibpunct{(}{)}{;}{a}{}{,} 

\usepackage{txfonts} 
\usepackage[dvipsnames]{xcolor}
\usepackage{booktabs} 
\usepackage{array} 
\usepackage{enumitem} 
\usepackage{amssymb} 
\usepackage{ulem} 

\definecolor{lime}{HTML}{A6CE39}
\DeclareRobustCommand{\orcidicon}{%
    \begin{tikzpicture}
    \draw[lime, fill=lime] (0,0) 
    circle [radius=0.16] 
    node[white] {{\fontfamily{qag}\selectfont \tiny ID}};
    \draw[white, fill=white] (-0.0625,0.095) 
    circle [radius=0.007];
    \end{tikzpicture}
    \hspace{-2mm}
}

\usepackage{tikz}

\foreach \x in {A, ..., Z}{%
    \expandafter\xdef\csname orcid\x\endcsname{\noexpand\href{https://orcid.org/\csname orcidauthor\x\endcsname}{\noexpand\orcidicon}}
}
\newcommand{\orcid}[1]{\href{https://orcid.org/#1}{\textcolor[HTML]{A6CE39}{\orcidicon}}}

\newcommand{\logP}{$\log{P}$}

\newcommand{\kms}{\,km\,s$^{-1}$} 
\newcommand{\ms}{\,m\,s$^{-1}$} 
\newcommand{\gaia}{\textit{Gaia}} %

\newcommand{\veloce}{\texttt{VELOCE}} %
\newcommand{\veloceDRI}{\texttt{VELOCE-I}} %
\newcommand{\nsample}{1000} 
\newcommand{\nsampledraws}{100} 

\begin{document}

\title{VELOCE III.\\Reconstructing Radial Velocity Curves of Classical Cepheids}
\author{
    Giordano Viviani \inst{1}\orcid{0009-0001-6201-2897}
    \and Richard~I. Anderson \inst{2,1}\orcid{0000-0001-8089-4419}\thanks{\email{richard.anderson@uni-goettingen.de}}
}

\authorrunning{G. Viviani \& R.I. Anderson}
\institute{
    Institute of Physics, \'Ecole Polytechnique F\'ed\'erale de Lausanne (EPFL), Observatoire de Sauverny, Chemin Pegasi 51b, 1290 Versoix, Switzerland
    \and
    Institut f\"ur Astrophysik und Geophysik, Georg-August-Universit\"at G\"ottingen, Friedrich-Hund-Platz 1, 37077 G\"ottingen, Germany\\
 }

\date{submitted: November 13, 2025; revised: September 24, 2026; accepted: September 25, 2026}

\abstract{We present a novel framework for accurately reconstructing radial velocity (RV) curves of classical Cepheids (Cepheids) from sparsely sampled time-series data suitable for application in large spectroscopic surveys. The framework provides a prior for the principal components of RV curves established based on high-precision measurements from the VELOcities of CEpheids (\veloce) project; template RV curves of Cepheids can be readily extracted from our results. We demonstrate the ability of our framework to estimate unbiased pulsation average velocities, $v_\gamma$, to within $20-30$\,\ms, and peak-to-peak amplitudes, $P2P$, to within $\sim 1.1\%$. Subsampling the initial data set, we show that $v_\gamma$ and $P2P$ can be determined to within $\sim 500$\,\ms and $\sim 4\%$, respectively, from as few as three observations drawn at random pulsation phases, with the phase of minimum radius fitted rather than assumed. As expected, the precision on both $v_\gamma$ and $P2P$ improves markedly with the number of observations, $N_{\mathrm{RV}}$; the precision on $P2P$ stabilizes beyond $N_{\mathrm{RV}}\simeq 10-15$, whereas that on $v_\gamma$ continues to improve. We fitted existing time-series RV data of Cepheids in the Large and Small Magellanic Clouds (LMC, SMC) using this framework and obtained typical root mean square errors of $0.5-2.0\,$\kms. The typical total uncertainty on $v_\gamma$ achieved for the SMC Cepheids is $\sim 0.84$\,\kms, providing sensitivity to spectroscopic binaries (SB). We identified 8 SB1 systems; two and one of which are new detections in the LMC and SMC, respectively. This yields a single-lined SB fraction of $\sim 25\%$ and $27\%$ in the two galaxies, similar to the Milky Way's SB fraction of $29\%$ established as part of \veloce. Despite their relatively small number, LMC Cepheids reproduce the known line-of-sight component of the LMC's large-scale rotation, which differs in the extremes by more than $80$\,\kms. The kinematics of the SMC are more complex and not sufficiently sampled by the available Cepheids. Our framework is designed to yield accurate $v_\gamma$ and $P2P$ of Cepheids observed by large spectroscopic surveys, such as 4MOST, SDSS-V, and others, and will unlock new insights into the kinematics and multiplicity of evolved intermediate-mass stellar populations.}


\maketitle
\section{Introduction}\label{sec:intro}
Classical Cepheids are cornerstone objects in astrophysics. Their radial pulsations cause periodic radius and temperature changes observable as variations in light intensity, angular diameter, and as spectral line shapes and positions. Radial velocity (RV) measurements based on high-resolution spectra encode fundamental stellar properties and are useful, among other things, for determining the multiplicity of evolved intermediate-mass stars \citep[cf.][for a review and \citealt{shetye_velocities_2024,Dinnbier2024} for Cepheids more specifically]{Moe2017}, calibrating and measuring distances via the Baade-Wesselink technique \citep[e.g.,][]{storm_bvrijk_2004, nardetto_high_2006, nardetto_high-resolution_2008, Anderson2014, Gallenne2017, Trahin2021}, and for understanding Galactic structure and kinematics \citep[e.g.,][]{1994A&A...285..415P,2019ApJ...870L..10M,2020ApJ...895L..12A}.

The advent of large time-resolved spectroscopic surveys enables the study of spectral variations in Cepheids in very large quantities, albeit typically with relatively few observations. A notable exception is the ESA {\it Gaia} mission \citep{prusti_gaia_2016}, which published time-series RV measurements based on the Ca~II IR triplet for 715 Galactic Cepheids as part of the third {\it Gaia} data release \citep[GDR3]{gaiadr3.dr3,gaiadr3.radvel,gaiadr3.cepheid}. 
Large spectroscopic surveys, notably the fifth installment of the Sloan Digital Sky Survey \citep[SDSS-V]{kollmeier_sdss-v_2017}, WEAVE \citep{WEAVE}, the 4-metre Multi-Object Spectroscopic Telescope \citep[4MOST]{4MOST} that recently saw first light, and future projects in development, such as the Wide-field Spectroscopic Telescope \citep[WST]{mainieri_wide-field_2024} can generally acquire a few epochs per object but are not designed for cadenced time-series observations. In particular, the One Thousand and One Magellanic Fields (1001MC) survey in 4MOST \citep{cioni_4most_2019} will collect high and low resolution spectra of $\sim 9'000$ Cepheids, with $3-6$ visits planned in the central regions and an expected per-epoch RV accuracy of $\sim 2$\kms. If accurate mean velocities of Cepheids can be established from such sparsely sampled RV curves, then extremely detailed line-of-sight velocity studies of the Magellanic Clouds \citep[cf.][]{vanderMarel2014,gaiadr3.lmckinematics} will come within reach and complement proper motion-based studies \citep[e.g.,][]{Niederhofer2022}. Additionally, this will provide insights into possible differences among binary fractions across environments, galaxies, and metallicity regimes.
Using Cepheids as precise tracers of kinematics requires considering their large-amplitude RV variations when determining the average, barycentric velocities, $v_\gamma$. Similar issues apply in the context of brightness variations in Cepheids \citep{Kanbur2002, Tanvir2005, Yoachim2009, Bhardwaj2017, bras_versatile_2025} and other pulsating stars, such as RR Lyrae stars \citep{Kanbur_Mariani_2004, Hajdu2018, Deb_Singh_2009, Paul_Chattopadhyay_2022}, where mean magnitudes are often determined from sparsely sampled light curves using template fitting. For RR Lyrae stars, several studies have furthermore constructed RV templates and analytic RV curve representations \citep[e.g.][]{Liu1991RRLRVtemplates, Sesar2012RRLRVtemplates, Braga2021, Huang2024, Prudil2024}. However, such an approach has thus far remained unfeasible in high detail for classical Cepheids due to the unavailability of sufficient data.

The high-quality homogeneous RV time series published in the first data release of the VELOcities of CEpheids project \citep[henceforth \veloceDRI]{anderson_velocities_2024}   provide a large and densely sampled reference data set for understanding Cepheid RV variations. In particular, \veloceDRI\ data have revealed new features of Cepheid RV curves (e.g., double bumps), allowed to determine the spectroscopic binary fraction of Galactic Cepheids \citep[henceforth \veloce-II]{shetye_velocities_2024} and to assess the quality of GDR3 RV time series of Cepheids, and enabled the detection of a large zoo of modulation signals, some of which are explained by non-radial pulsations \citep{Anderson2014, anderson_discovery_2016, Anderson2019, anderson_velocities_2024, netzel_veloce_2024, netzel_2025_velocemodzoo2, barbey_veloce_2025}.

Here, we employed Principal Component Analysis (PCA) to establish the first high-fidelity framework  for reconstructing classical Cepheid RV curves  based on \veloceDRI.
The PCA method is a dimensionality reduction technique frequently used in the creation of light and RV curve templates. It identifies a set of successive orthogonal components, called principal components (PCs), that sequentially maximize the explained variance within a given dataset of correlated variables \citep{Pearson1901, Hotelling1933}. Thus, PCA aims to pinpoint a subset of dimensions that can best re-express the given dataset as linear combination of the PCs \citep{Shlens2014_pca_linear}. In practice, it allows to represent RV curves by a combination of projection coefficients. Templates are usually constructed by determining how these coefficients vary with a chosen set of physical parameters. This mapping, however, is often not unique. In these cases, templates are not always reliable and it is better to infer probability distributions of the coefficients. These can be used as priors when data are available, and can generate tentative templates based on the most probable coefficients.
This PCA-based framework for the reconstruction of Cepheid RV curves is useful for investigating the structure and kinematics of the Milky Way and resolved galaxies, as well as for detecting spectroscopic binaries in great numbers and across a variety of environments based on relatively few observations; especially in the future as spectroscopic surveys will increasingly incorporate the temporal dimension into their strategies \citep{mainieri_wide-field_2024}.

Measuring galaxy kinematics and structure implies that observed samples differ substantially in metallicity, both within the Milky Way, where spectroscopic surveys have established an abundance range of about 1\,dex across the Galactic disc \citep[e.g.,][]{luck_distribution_2011, genovali_fine_2014, trentin_cepheid_2023, da_Silva_2023, Trentin_2024}, and between galaxies, the mean metallicities of Cepheids in the Large and Small Magellanic Clouds (LMC, SMC) being significantly lower than Solar \citep{romaniello_iron_2022, breuval_small_2024}. It is therefore necessary to assess how RV curve templates calibrated on Galactic Cepheids of approximately Solar metallicity perform when applied to more metal-poor populations. Metallicity may plausibly affect RV curve morphology through its influence on the pulsation itself: opacities depend on chemical composition, and the resonance responsible for the Hertzsprung progression, which shapes the bump features of both light and RV curves, is sensitive to the stellar structure. Empirically, metallicity has been reported to affect light curve shapes \citep[e.g.][]{klagyivik_determination_2013, hocde_metallicity_2023} and the projection factor for BW analyses \citep{nardetto_baade-wesselink_2011}. Studies focused on RVs have reported mixed results: early works found evidence for larger RV amplitudes at lower metallicity \citep{pont_metallicity_2001, klagyivik_study_2007, szabados_observational_2012}, whereas \citet{hocde_precise_2024} found no such correlation and called for further investigation.

This paper is organized as follows. Section~\ref{sec:data_pca} describes the data and application of the PCA to determine the PCs, and evaluate their overall performance. Section~\ref{sec:prior_nrv_fitting} presents the methodology used to determine prior probability distributions for the PC coefficients, defines our framework for fitting sparsely sampled time-series, and evaluates its precision and limitations.
Section~\ref{sec:extragal} reports the methodology and results of the application of our framework to literature RV measurements of Cepheids in the LMC and SMC. The final Section~\ref{sec:conclusion} summarizes our results and presents our conclusions and perspectives.

\section{Data and Principal Component Analysis}\label{sec:data_pca}

This work adopted the following nomenclature to distinguish between related, yet conceptually distinct, quantities related to RV analysis. Radial velocities measured from spectroscopic observations are referred to as RV measurements. The intrinsic RV variation of a Cepheid over its pulsation cycle is referred to as its RV curve, and we consider the Fourier series models ($\texttt{V}^{\rm FS}$), which were determined and carefully checked as part of \veloceDRI, as the corresponding ground truth. Continuous functions used to represent the pulsational RV curve based on the PCs are referred to as $\mathcal{M}$. Unless otherwise noted, $\mathcal{M}$ is established by fitting PCs to data points drawn from the $\texttt{V}^{\rm FS}$ models. A distinction is made for PC-based models fitted to RV measurements, which are labeled $\mathcal{M}_{\rm RV}$.

Using this nomenclature, Section \ref{subsec:data} provides an overview of the data used in this work. Section \ref{subsec:pca_determination} describes the methodology and outlines the determination of the PCs, while Section \ref{subsec:pca_evaluation} presents the evaluation of their performance over the \veloceDRI\ sample.

\begin{figure*}[th]
    \centering
    \includegraphics[width=0.85\textwidth]{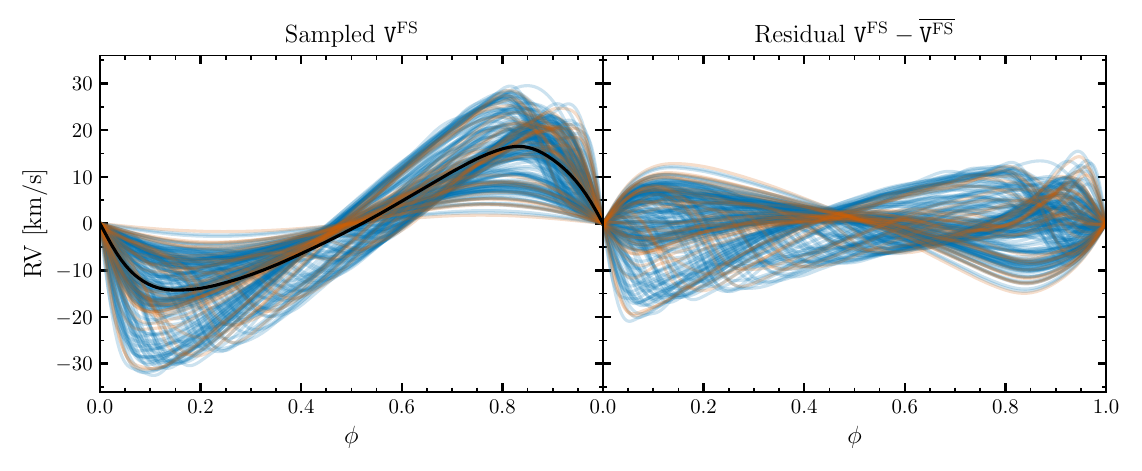}
    \caption{\label{fig:rv_curves}Cepheids' RV curves published in \veloceDRI. On the left panel, we report the sampled $\texttt{V}^{\rm FS}$ for both the training (blue) and test (orange) set. The solid black line indicates the mean curve of the training set, $\overline{\texttt{V}^{\rm FS}}$. The residual curves after subtracting $\overline{\texttt{V}^{\rm FS}}$ are reported in the second panel.}
\end{figure*}

\subsection{\veloce\ observations and FS models\label{subsec:data}}

\veloceDRI\ published more than $18\,000$ RV measurements of $258$ Galactic classical Cepheids across both hemispheres. High-precision RV measurements were obtained via the cross-correlation technique \citep{Baranne1996}, resulting in typical uncertainties ranging from $2$ to $50$ \ms, with a median of 37\ms. The extensive dataset yields well-sampled pulsation RV curves over a decade-long baseline for most targets, enabling the characterization of the pulsation curve for $219$ Cepheids. In \veloceDRI, pulsational variability was modeled using Fourier series (FS), supplemented by a polynomial for cases exhibiting a variable average velocity, e.g., in the case of suspected binaries. Where necessary, we isolated the pulsational variability from the RV measurements by subtracting variations in $v_\gamma$ using these polynomial trends.
We adopted the ephemerides from \veloceDRI\ to ensure consistent alignment of all RV curves. Following \cite{anderson_discovery_2016}, pulsation phase, $\phi$, is defined such that $\phi=0$ corresponds to minimum radius, i.e., the intersection of the descending branch with $v_\gamma$. This convention optimizes precision for $\phi$ and differs from the definition generally used in light curve analyses, where $\phi=0$ usually refers to maximum light, which typically occurs $\Delta \phi \approx 0.1-0.2$ after minimum radius.

\subsection{Principal Component Analysis}
\label{subsec:pca_determination}

We applied PCA to finely sampled pulsational RV models of 219 Cepheids constructed by sampling the \veloceDRI\ Fourier series models on a uniform grid of $N_{dim}=$\nsample\ points between $0\leq \phi < 1$ \citep[as done, e.g., in RR Lyrae light curves by][]{Hajdu2018}. Since $\texttt{V}^{\rm FS}$ were carefully determined in \veloceDRI, we considered them a satisfactory approximation to the true RV curve. 
Previous PCA analyses of light curves have generally been performed on FS coefficients rather than on the curves themselves \citep[e.g.][]{Kanbur2002, Kanbur_Mariani_2004, Tanvir2005, Yoachim2009}, typically adopting the linear formulation of the FS, for which curves fitted with different numbers of harmonics can be accommodated by setting the unused coefficients to zero, provided that the coefficients are not subsequently standardized, which would otherwise inflate the variance associated with the rarely fitted higher harmonics. We nonetheless preferred to work directly with discrete model points, since this representation is independent of the parametric form adopted and of the harmonic order fitted to each star (which varies with the sampling and complexity of each individual RV curve) and places all targets on a common phase grid, so that the PCA operates on curve shape alone. All sampled curves were assigned identical weights for simplicity.

The dataset was divided into training and test sets by randomly setting aside 15\% of the targets for testing. After examining several normalization strategies, we decided to remove the mean value of the training $\texttt{V}^{\rm FS}$ at each point of the grid, $\overline{\rm{\texttt{V}^{FS}}}$, without applying any kind of standardization to the single curves and/or along each dimension (phase bin). This choice ensures that the amplitude information of the curves is maintained in the residuals, $\rm{\texttt{V}^{FS}} - \overline{\rm{\texttt{V}^{FS}}}$, and can be captured by the PCs. Figure\,\ref{fig:rv_curves}\ shows the $\texttt{V}^{\rm FS}$ used before and after this process. 

We performed PCA of the training set using the \texttt{scikit-learn} Python library \citep{pedregosa_scikit-learn_2011} and obtained the individual PCs which relate to the original curves as follows:
\begin{equation}
    \label{eq:PCrelation}
    \mathrm{\texttt{V}^{FS, t}}(\phi) = \overline{\rm{\texttt{V}^{FS}}}(\phi) + \sum^{N_{dim}}_{i=1} p_i^{t} \ \mathrm{PC}_i(\phi) \ ,
\end{equation}
where $p_i^{t}$ are the projection coefficients for a specific target $t$.

\begin{figure}
    \centering
    \includegraphics[width=0.4\textwidth]{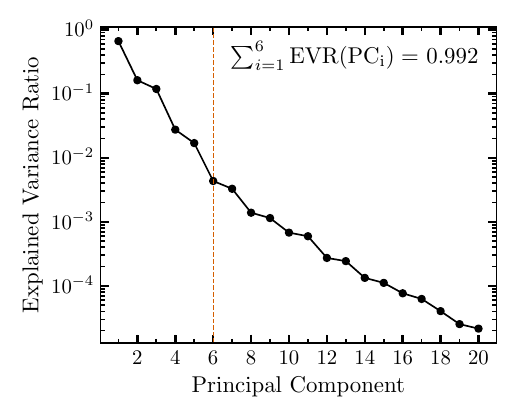}
    \caption{\label{fig:pca_variance_ratio}Variance ratio explained (in log scale) by each PC. The vertical orange dashed line separates the retained (left) and discarded (right) PCs. The total explained variance ratio of the retained PCs is shown in right top corner.}
\end{figure}

\begin{figure}[hbt!]
    \centering
    \includegraphics[width=0.48\textwidth]{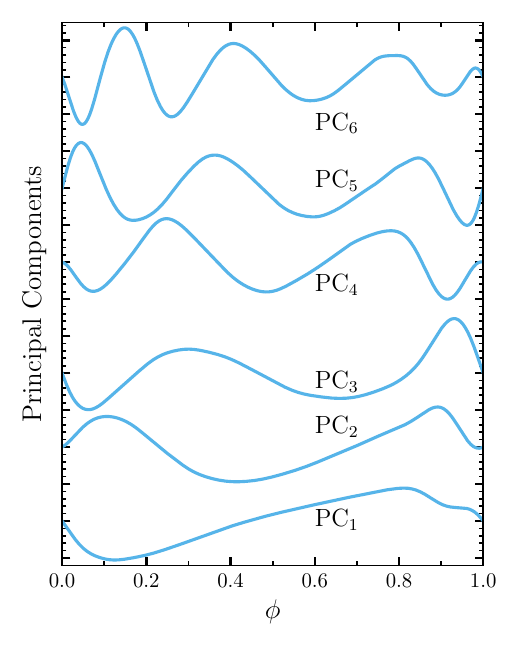}
    \caption{\label{fig:pca_components} From bottom to top: the 6 PCs obtained by decomposing the \veloceDRI\ FS models, $\texttt{V}^{\rm FS}$.}
\end{figure}

We determined the optimal number of PCs ($N_{\rm PC}$) to retain by analyzing the explained variance, selecting the smallest number of components that capture a sufficiently high fraction of the total variance, thus balancing dimensionality reduction and reconstruction accuracy. The explained variance-ratio (EVR) obtained for each of the first 20 PCs is shown in Figure \ref{fig:pca_variance_ratio}. Most of the variance is captured by the first few components, as seen in PCA of Cepheid light curves \citep[e.g.][]{bras_versatile_2025}. 
This behavior persists even when PCA is performed separately on the fundamental-mode and overtone subsamples, indicating no significant mode-dependent reduction in dimensionality. We therefore chose not to partition the training sample further and retain the first 6 components ($N_{PC}=6$), which cumulatively explain $\sim 99.2\%$ of the total variance. The corresponding PCs are displayed in Fig.\,\ref{fig:pca_components}.

\begin{figure}
    \centering
    \includegraphics[width=0.5\textwidth]{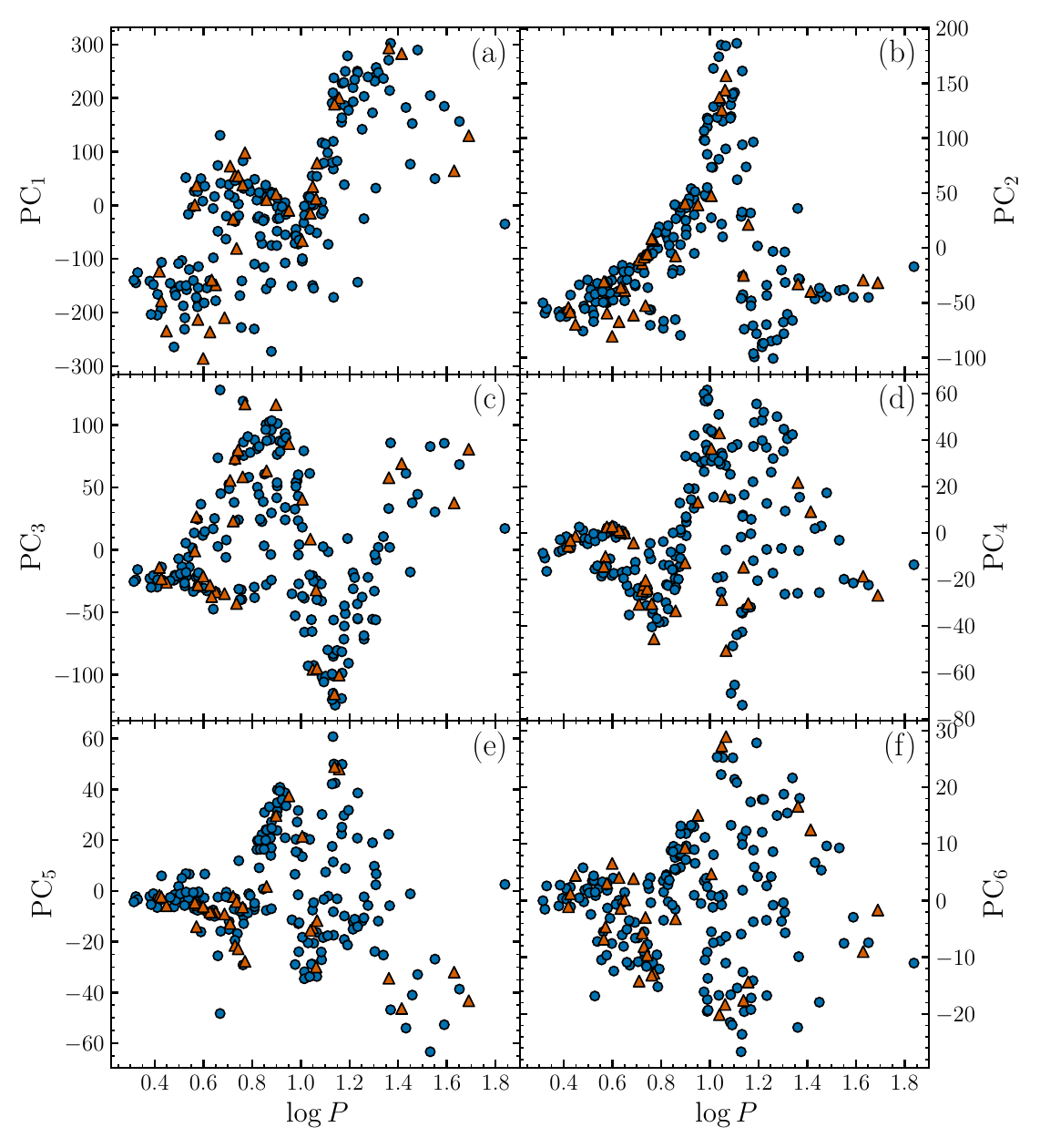}
    \caption{\label{fig:pc_FS_distribution} Distribution of the coefficients $p_i$ of the $\mathcal{M}$ models as a function of the logarithm of the pulsation period, \logP. As for the previous plots, the training set is presented in blue, whereas the test set in orange. The values obtained for the outliers discussed in Section\,\ref{subsec:pca_evaluation} are not shown.}
\end{figure}

Using these PCs, we fitted the $\texttt{V}^{\rm FS}$ models of both training and test sets to obtain the reconstructed models, $\mathcal{M}$, and  their respective coefficients $p_i$.
Figure\,\ref{fig:pc_FS_distribution} shows the distribution of the coefficients $p_i$ against \logP. The coefficients for training and test targets follow consistent trends. Interestingly, the PCs bear some resemblance to the distributions of Fourier parameters shown in Fig.\,11 of \veloceDRI. $\rm{PC_{1}}$ resembles the distribution of peak-to-peak amplitudes, indicating that the first component primarily accounts for the variability associated with the overall amplitude of the pulsation. The distributions of $\rm{PC_{2}}$ and $\rm{PC_{3}}$ mostly resemble the Fourier amplitude ratios, $R_{21}$ and $R_{31}$, albeit with some differences and some possible resemblance between $\rm{PC_{2}}$ and the Fourier phase difference $\phi_{21}$ as well.
Higher-order PCs exhibit increasingly less obvious trends with \logP, as is also the case for the higher-order Fourier amplitude ratios, whereas the higher-order Fourier phase differences retain well-defined progressions (\veloceDRI, Fig. 11 and Appendix A.1).

\subsection{Evaluating PC Performance}
\label{subsec:pca_evaluation}

\begin{figure*}
    \centering
    \includegraphics[width=.85\textwidth]{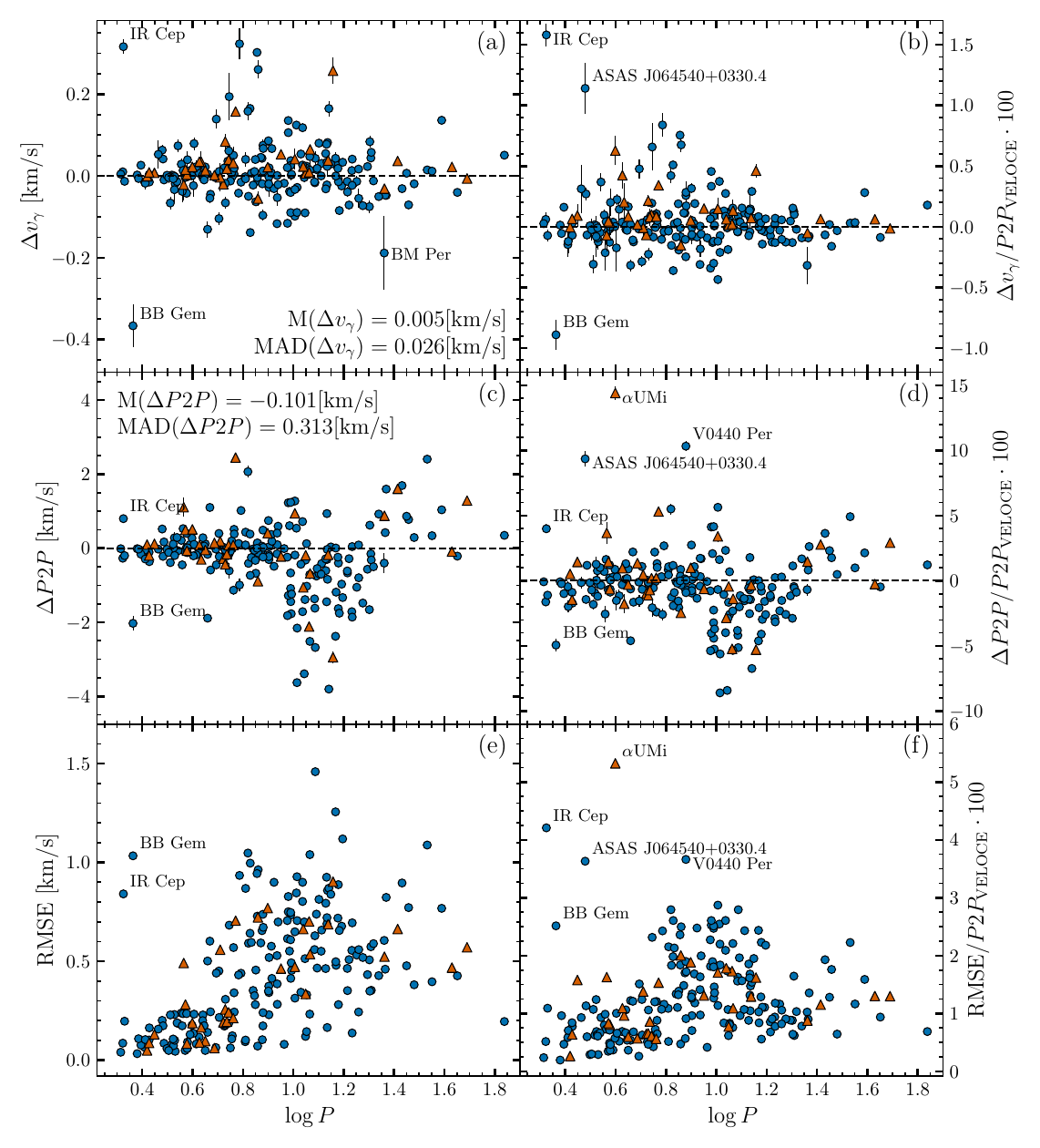}
    \caption{\label{fig:results_fit_rv_stats}Performance of the PCs when fitting \veloceDRI\ RV measurements. The training and test set members are colored in blue and orange respectively. The left panels show, from top to bottom, the difference in pulsation-averaged velocity, $\Delta v_\gamma$, the difference in peak-to-peak amplitude, $\Delta P2P$, and $RMSE$. The right panels present the same quantities as a percentage of the target's $P2P_{\rm VELOCE}$. Panels (a) and (c) report the median, M, and the median absolute deviation, MAD, of the corresponding distributions. AV\,Sgr is not visible, as its values fall well outside the range shown. }
\end{figure*}

To evaluate the ability of the retained PCs to reconstruct Cepheids' RV curves from observations alone, we fitted the \veloce\ RV measurements and compared the results with the FS models published in \veloceDRI, $\texttt{V}^{\rm FS}$. The fit was performed via least-squares optimization, linearly interpolating the PCs between the grid points\footnote{For the interpolation, we used the \texttt{numpy.interp} function of the \texttt{NumPy} Python library \citep{harris_array_2020}}. As quantitative metrics, we considered the pulsation-averaged velocity, $v_\gamma$, and the peak-to-peak amplitude, $P2P$, comparing both with the \veloce\ results. We also computed the root-mean-squared error, $RMSE$, between the fitted PC model, $\mathcal{M}_{\rm RV}$, and $\texttt{V}^{\rm FS}$, each evaluated at 1000 equidistant values of $\phi$, to assess the average discrepancy between the two. Additionally, we normalized each of these metrics by the $P2P$ amplitude of each star as given by $\texttt{V}^{\rm FS}$, $P2P_{\mathrm{VELOCE}}$, expressing them as percentages, since amplitudes depend strongly on \logP\ and pulsation mode. Figure\,\ref{fig:results_fit_rv_stats} illustrates these quantities as a function of $\log P$, while Figure\,\ref{fig:example_fit_rv} shows the comparison between $\texttt{V}^{\rm FS}$ and $\mathcal{M}_{\rm RV}$ for a few targets of the training and test sets. 

Training and test sets exhibit consistent behaviors throughout the entire period range, showing no discrepancies that would indicate overfitting, an unfavorable dataset split, or other limitations of the analysis. Hence, the PCs appear to generalize well to unseen data. 

We identified six Cepheids as outliers: AV\,Sgr, BB\,Gem, IR\,Cep, $\alpha$\,UMi (Polaris), V0440\,Per, and ASAS\,J064540+0330.4. The first three have unreliable fits and were removed from all subsequent analysis. AV\,Sgr's nine RV measurements all fall on the ascending branch of the pulsation, thereby poorly constraining any RV curve fit. For BB\,Gem and IR\,Cep, the sparsely sampled maximum of the RV curve and the complex structure of the data, respectively, yielded fitted PC models that deviated from the corresponding FS models; in both cases the two solutions reproduce the RV measurements comparably well, with no objective criterion favoring one over the other. The outlier nature of the remaining three stars, $\alpha$\,UMi, V0440\,Per and ASAS\,J064540+0330.4, is explained by their unusually low amplitudes for their respective \logP, which were not sufficiently represented in the training data. As such stars will inevitably be encountered when applying our framework to external data sets, we retained these targets in our analysis.\\

\subsubsection{Pulsation-averaged velocity}\label{subsubsec:vgamma_all}

Fig.\,\ref{fig:results_fit_rv_stats}a shows the difference
$\Delta v_\gamma = v_\gamma - v_{\gamma, \rm VELOCE}$ for each Cepheid under study and indicates strong agreement between the pulsation-averaged velocities of the $\mathcal{M}_{\rm RV}$ and $\texttt{V}^{\rm FS}$ models.
The scatter in $\Delta v_\gamma$ is centered around $0$ across \logP, with a mean of $11$\ms\ and a standard deviation of $66$\ms. The corresponding median and median absolute deviation (MAD), less affected by the few outlying targets, are 5 and 26\ms.
The training and test samples show consistent results with mean (standard deviation) values of $9$\ms\ ($67$\ms) and $28$\ms\ ($55$\ms), respectively, indicating that the obtained PCs generalize well. As mentioned above, BB\,Gem and IR\,Cep appear as outliers for \logP $<0.4$, while AV\,Sgr falls beyond the range shown. The large error for BM\,Per is due to an uneven sampling of RV measurements mostly located on the ascending branch of the RV curve, similarly to AV\,Sgr. 

The ratio $\Delta v_\gamma/P2P_{\mathrm{VELOCE}}$, which removes the effect of the large differences in amplitude among stars, is shown in Fig.\,\ref{fig:results_fit_rv_stats}b. For $90\%$ of the targets, $v_\gamma$ is recovered within $0.31\%$ of $P2P_{\mathrm{VELOCE}}$.
The scatter is particularly small for $\log{P} > 1$, and overall appears to decrease with \logP, and thus generally the amplitude.

\subsubsection{Peak-to-peak amplitude}\label{subsubsec:p2p_all}

Figure\,\ref{fig:results_fit_rv_stats}c illustrates the results for $\Delta P2P = P2P - P2P_{\rm VELOCE}$ and Fig.\,\ref{fig:results_fit_rv_stats}d the corresponding relative differences. The relative differences exhibit a MAD of $1.1\%$, and 90\% of the amplitudes are recovered to within $4.1\%$ of the expected values, or $\sim1.6$\kms\ in absolute terms. As mentioned before, stars with anomalously low amplitudes ($\alpha$\,UMi, V0440\,Per and ASAS\,J064540+0330.4) appear as clear outliers. 
Both panels reveal a tendency to systematically underestimate amplitudes between $9.7 < P \mathrm{[d]} < 20$. Below $P \sim 9.7\,$d ($\log P \sim 0.987$), both $\Delta P2P$ and $\Delta P2P/P2P_{\mathrm{VELOCE}}$ are symmetric about $0$, and their scatter remains narrow and nearly constant, with $90\%$ of the values within merely $0.74$\kms. At longer periods, the fact that the Hertzsprung progression (HP) is not a unique function of \logP\  \citep[][Figs.\,11 and 12 in \veloceDRI]{hertzsprung_relation_1926, soszynski_optical_2008} introduces an astrophysical limitation to how closely RV curve shapes can be generalized: $P2P$ is typically somewhat underestimated for \logP $\in [1,1.3]$, by 1.5\% in the median, and slightly overestimated at \logP$> 1.3$. The $\mathcal{M}_{\rm RV}$ models of FN\,Aql, V2475\,Cyg, AD\,Pup, and X\,Pup in Figure\,\ref{fig:example_fit_rv} clearly illustrate the limitations of the PCs in correctly reconstructing the RV curves near the HP bump. For very long-period Cepheids, further limitations include the lower density of stars (in \logP) and the difficulty of achieving sufficiently good phase coverage for stars exhibiting significant cycle-to-cycle variations and complex period changes \citep[\veloceDRI]{Anderson2014,csoernyei_2022}.

\subsubsection{Overall shape of the RV curve\label{sec:rmse}} 
The $RMSE$ computed between $\mathcal{M}_{\rm RV}$ and $\mathrm{\texttt{V}^{FS}}$ quantifies the average difference between the two models in \kms. Figures\,\ref{fig:results_fit_rv_stats}e and \ref{fig:results_fit_rv_stats}f show the $RMSE$ in \kms\ and as a percentage of $P2P_{\rm VELOCE}$, respectively. Overall, these distributions confirm the strong performance of the PCs in accurately reconstructing the $\mathrm{\texttt{V}^{FS}}$ curves from the RV measurements.
The mean $RMSE$ and its scatter both increase with the pulsation period. For most Cepheids with $\log P< 0.8$, $RMSE$ remains below $0.3$\kms, while for longer-period stars, which are characterized by the HP bump, the threshold rises to $\sim 1$\kms. The 90th percentile of the distributions is found at $0.83$\kms, and at $\sim2.2$\% of the total pulsation amplitude, while the median values are $0.42$\kms and $\sim1.1$\%. The relative $RMSE$ distribution in Fig.\,\ref{fig:results_fit_rv_stats}f shows a weaker dependence on the pulsation period, yet confirms that the PCs cannot reliably reproduce the HP bump for Cepheids within $\log P \in [1,1.3]$. The envelope of lowest relative $RMSE$ in Fig.\,\ref{fig:results_fit_rv_stats}f also exhibits a clear trend with \logP. As expected, the outliers discussed above all exhibit larger $RMSE$ values than the rest of the targets.

 \section{Priors determination and RV curve fitting}\label{sec:prior_nrv_fitting}
A key application of the PCs calibrated in this work lies in the determination of RV curves from sparsely sampled time series with particular emphasis on accurately estimating $v_\gamma$ in order to investigate stellar multiplicity and galaxy kinematics. The small number of retained PCs, combined with the information encoded in both the components and their coefficients, can be exploited to reconstruct the RV curve from sparse measurements more reliably than is possible with general models, such as Fourier series, which incorporate no prior knowledge of the Cepheid RV curve morphology.
 
Section\,\ref{subsec:prior_pcs} and Section\,\ref{subsec:prior_delta_phi} present the determination of prior probability distributions for the PC coefficients and for phase correction, $\Delta\phi$, that aligns the reference epoch with minimum radius (see Section\,\ref{subsec:data}). Section\,\ref{subsec:methodology_map} describes the methodology adopted to fit a varying number of RV measurements, allowing us to evaluate the performance of the framework, notably in the regime of few available observations. Finally, Section\,\ref{subsec:result_framework} presents and discusses the results obtained.
 
 \subsection{Prior probability on PC coefficients \label{subsec:prior_pcs}}
 
\begin{figure*}[th!]
    \centering
    \includegraphics[width=.95\textwidth]{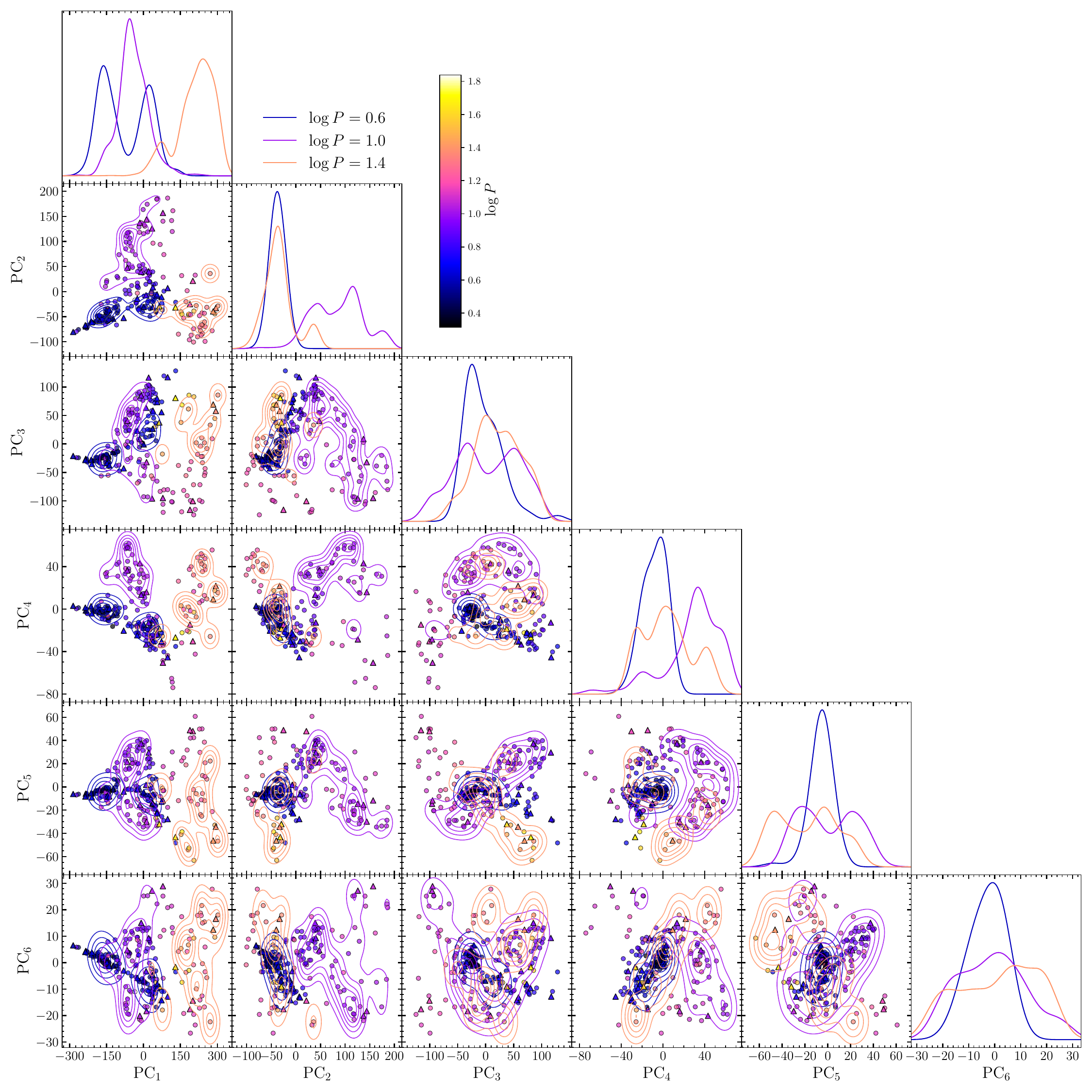}
    \caption{Corner plot of the joint probability distribution of the six principal-component coefficients $p_i$ used to reconstruct radial-velocity curves, estimated from the training set via a multivariate KDE described in Section\,\ref{subsec:prior_pcs}. Diagonal panels show the one-dimensional marginal conditional density $\mathcal{P}(p_i \mid \log P)$, obtained by conditioning on the pulsation period and marginalizing over all remaining coefficients, for three representative periods: \logP\ $= 0.6, 1.0$, and $1.4$ (short, medium, and long; color-coded as in the colorbar). Off-diagonal panels show density contours of the two-dimensional conditional distribution $\mathcal{P}(p_i, p_j \mid \log P)$ at the same three periods. Coefficients of the $\mathcal{M}$ fits from the training (circles) and test (triangles) sets are overlaid, with colors indicating their pulsation period. \label{fig:marginal_kde}}
\end{figure*}

As discussed in Section\,\ref{subsec:pca_determination}, the PC coefficients $p_i$ $(i =1...N_{\rm PC})$ describe the morphology of a Cepheid RV curve, and their distributions as a function of \logP\ exhibit strong trends reminiscent of the Fourier amplitude ratios and the peak-to-peak amplitude. Their joint density therefore encodes substantial prior knowledge of the RV curve shape, which can be exploited to constrain reconstructions when few observations are available. Moreover, the coefficients are not mutually independent: the orthogonality of the PCs does not imply statistical independence of the coefficients, and we find residual correlations among $p_i$, in particular at fixed pulsational period (cf. Fig.\,\ref{fig:marginal_kde}). Capturing these correlations, rather than treating each coefficient in isolation, yields tighter and more physically faithful priors.
For this reason, we estimated the density distribution of the full seven-dimensional space (\logP, $p_1, p_2, ..., p_6$) from the training set, rather than modeling each coefficient separately. We used the multivariate kernel density estimation (KDE), \texttt{KDEMultivariate}, implemented in the \texttt{statsmodels} Python library \citep{seabold2010statsmodels}, with the bandwidth in each dimension optimized via the built-in cross-validated maximum likelihood selector (\texttt{bw=cv\_ml}).
Conditioning the resulting empirical distribution at a given \logP\ yields the prior probability distribution of the coefficients for a Cepheid of that period, $\mathcal{P}(p_1, p_2, \ldots, p_6 \mid \log P)$. Estimating the density jointly thus captures both the dependence of the coefficients on \logP\ (cf. Fig.\,\ref{fig:pc_FS_distribution}) and the correlations among them, which constitutes a more general treatment of these correlations than, for example, the separate priors on ratios of PC coefficients adopted by \citet{Hajdu2018}.

Two practical considerations apply. First, the KDE becomes unreliable beyond the ranges spanned by the training set, most notably in \logP. Similarly to \citet{Hajdu2018}, we restricted the prior to the range of each dimension padded by 10\% of its total extent. Second, the resulting seven-dimensional distribution and its conditioning on \logP\ cannot be visualized directly. In Fig.\,\ref{fig:marginal_kde}, we therefore show a corner plot of the six coefficients, presenting the joint conditional distributions $\mathcal{P}(p_i, p_j \mid \log P)$ together with the marginal conditional distributions $\mathcal{P}(p_i \mid \log P)$ at selected values of \logP.

\subsection{Prior probability on $\Delta\phi$ from photometry\label{subsec:prior_delta_phi}}
While the pulsation period of a Cepheid targeted for RV observations can generally be assumed known, the phase of the RV measurements relative to minimum radius ($\phi = 0$ in our convention) is not. Fitting the RV measurements of a given Cepheid therefore requires an additional free parameter, $\Delta\phi$, that aligns the reference epoch adopted for phase-folding with minimum radius (as is also done in our application to extragalactic Cepheids; cf. Section\,\ref{subsec:method_MC}). Constraining $\Delta\phi$ with an informative prior is particularly valuable in the context of few available observations, where it would otherwise be poorly determined.

Such a prior can be constructed from photometry, under two assumptions that are well justified in practice. First, RV observations are obtained contemporaneously with, or subsequent to, photometric measurements, so that a light-curve reference epoch is available for the target. Second, light curves are conventionally phased to maximum light, which follows minimum radius by roughly $\Delta\phi \simeq 0.1-0.2$. The offset between the photometric reference epoch and minimum radius therefore carries direct prior information on $\Delta\phi$.

We calibrated this prior using \gaia\,DR3 \citep{gaiadr3.dr3, gaiadr3.cepheid}, whose photometric time series are fully contemporaneous with \veloce. The photometric and RV reference epochs are consequently close in time for most targets, minimizing complications from variable ephemerides or orbital motion. For each Cepheid we computed:
\begin{equation}
\Delta\phi = \left[\left(\frac{E_{\rm Gaia\,DR3} - E_{\rm VELOCE}}{P_{\rm VELOCE}} + 0.5\right) \bmod 1\right] - 0.5\,,
\label{eq:real_phase_diff}
\end{equation}
mapped onto $[-0.5, 0.5)$, where $E_{\rm Gaia\,DR3}$ is the \gaia\ DR3 lightcurve reference epoch in the G band and $E_{\rm VELOCE}$ and $P_{\rm VELOCE}$ are the reference epoch and period from \veloceDRI. Because $E_{\rm VELOCE}$ corresponds to minimum radius (see Section\,\ref{subsec:data}), $\Delta\phi$ measures the offset between maximum light and minimum radius directly. Extending the procedure to other photometric bands would follow identically.

Although \veloce\ and \gaia\,DR3 data sets overlap in time, both span long baselines, and observations of a given target may be separated by many pulsation cycles. Over such intervals, even small differences between the \gaia\ and VELOCE periods, due to intrinsic period changes or measurement uncertainties, can accumulate into a significant drift in $\Delta\phi$. We quantified this by computing the number of elapsed cycles, $N_{\rm cycles} = \left|E_{\rm Gaia\,DR3} - E_{\rm VELOCE}\right| / P_{\rm VELOCE}$, and the corresponding phase drift,
\begin{equation}
\delta_{\Delta\phi} = \left|N_{\rm cycles}\cdot\left(\frac{P_{\rm Gaia\,DR3} - P_{\rm VELOCE}}{P_{\rm VELOCE}}\right)\right|\,.
\end{equation}
Targets with $\delta_{\Delta\phi} > 0.1$ were considered unreliable and excluded.

\begin{figure}[th!]
    \centering
    \includegraphics[width=.49\textwidth]{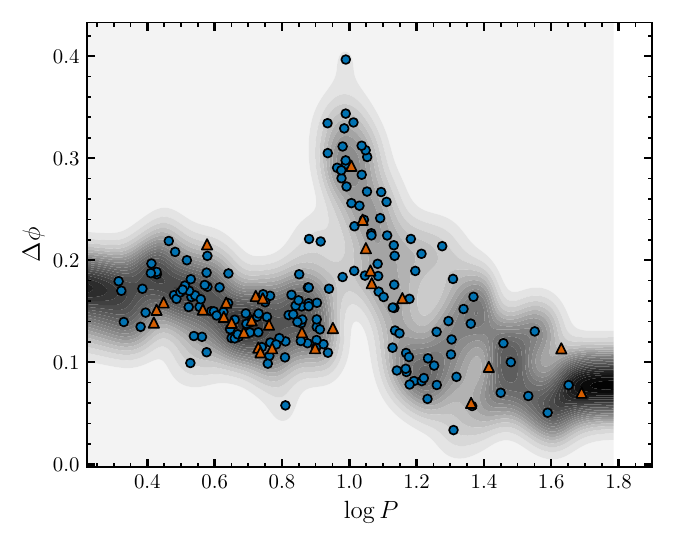}
    \caption{Phase offset $\Delta\phi$ between maximum light and minimum radius as a function of $\log P$. As in previous plots, the training and test sets are shown in blue and orange, respectively. The KDE distribution estimated from the training set is shown in grayscale in the background. For clarity, its density has been normalized to unit integral within each \logP\ slice (i.e. vertical slice).   \label{fig:phase_prior}}
\end{figure}

Of the 216 \veloce\ targets analyzed, 11 are absent from the \texttt{gaiadr3.vari\_cepheid} table (BB\,Sgr, FM\,Aql, RT\,Aur, SZ\,Aql, T\,Mon, TT\,Aql, TZ\,Mon, U\,Sgr, $\alpha$\,UMi, $\beta$\,Dor, $\zeta$\,Gem) and a further 4 exceed the drift threshold (ASAS\,J190929+1232.8, S\,Vul, V0389\,Sct, V1197\,Cen), leaving 201 reliable $\Delta\phi$ values, shown in Figure\,\ref{fig:phase_prior}.
As for the PC coefficients, we estimated the prior $\mathcal{P}(\Delta\phi \mid \log P)$ by applying a two-dimensional KDE to the (\logP, $\Delta\phi$) distribution and conditioning on \logP\ (using the same \texttt{statsmodels} implementation and  \texttt{cv\_ml} bandwidth selection as done in Section\,\ref{subsec:prior_pcs}). As Figure\,\ref{fig:phase_prior} shows, this prior is truncated at a shorter maximum period ($\max (\log P) \simeq 1.79$) than the PC coefficients prior, since \veloce's longest period Cepheid, S\,Vul, lacks a reliable $\Delta\phi$ and has been excluded.

\subsection{Application of the framework to sparse RV measurements \label{subsec:methodology_map}}

The PCs obtained in Section\,\ref{subsec:pca_determination} combined with the priors derived in Sections\,\ref{subsec:prior_pcs} and \ref{subsec:prior_delta_phi} define a framework whose key application is the reconstruction of RV curves from few measurements, as is typical of large spectroscopic surveys. To evaluate the performance of our framework in this regime, we simulated sparsely sampled observations by drawing random subsets of $N_{\rm RV}$ measurements from the VELOCE-I data, reconstructing the RV curve from each subset alone. We considered $N_{\rm RV} \in [3, 4, 5, 6, 7, 10, 15, 20]$ and, for each Cepheid and each $N_{\rm RV}$, drew \nsampledraws\ subsets without replacement, fitting each independently. Drawing many subsets characterizes the distribution of fits arising from the random choice of observed epochs, thus the sampling variability intrinsic to a sparse observing strategy, rather than the posterior uncertainty of any single data set. To limit correlation among subsamples, targets with $N_{\rm RV} + 2$ or fewer available measurements were excluded; the sole exception was V1197\,Cen (9 observations), for which the 84 distinct combinations available for $N_{\rm RV} = 3$ and 6 were used.

To reproduce the realistic situation of an external survey for which the Cepheids' phase relative to minimum radius is not known a priori, we phased the measurements on a photometric time reference and fitted the offset to minimum radius, $\Delta\phi$, as a free parameter. The phasing reference, $E_G$, was set to the \gaia\,DR3  G-band epoch for the 205 targets for which it was available (see Section\,\ref{subsec:prior_delta_phi}); for the remaining targets, we instead adopted $E_G = 2455197.5$ (=J2010.0), the epoch offset used in the \texttt{gaiadr3.vari\_cepheid} table \citep{gaiadr3.cepheid}. The measurements were then phased as $\phi_i = \left[(\mathrm{BJD}_i - E_G)/P_{\rm VELOCE}\right] \bmod 1$, and the PC-based model evaluated as
\begin{equation}\label{eq:map_model}
\mathcal{M}_{\rm RV}(\phi;\boldsymbol{\theta}) = v_\gamma + \overline{V^{\rm FS}}(\phi + \Delta\phi) + \sum_{i=1}^{6} p_i\,\mathrm{PC}_i(\phi + \Delta\phi)\,,
\end{equation}
where $v_\gamma$ is the systemic velocity, $p_i$ the six PC coefficients, and $\mathrm{PC}_i$ and $\overline{V^{\rm FS}}$ the principal components and training-set mean curve of Eq.\,\ref{eq:PCrelation}. The parameter vector is thus $\boldsymbol{\theta} = (v_\gamma, p_1, \ldots, p_6, \Delta\phi)$, comprising eight free parameters. 
This model was fitted to each subset with a Maximum a Posteriori (MAP) estimator. MAP estimation is the analog of the more widely used Maximum Likelihood estimation (MLE) in which the maximized quantity is a posterior rather than a likelihood \citep{bassett_maximum_2019}, and it can equivalently be regarded as a regularized form of MLE. Conceptually, MAP refines the parameter estimates by balancing the information contained in the data (as in MLE) with prior knowledge of the parameters. In the regime of few observations, where the data alone constrain the RV curve only weakly, this prior knowledge becomes decisive.

Assuming Gaussian residuals, the log-likelihood is
\begin{equation}\label{eq:map_likelihood}
\ln \mathcal{L}(\boldsymbol{\theta}) = -\frac{1}{2}\sum_{i=1}^{N_{\rm RV}} \left(\frac{v_i - \mathcal{M}_{\rm RV}(\phi_i;\boldsymbol{\theta})}{\sigma_{{\rm eff},i}}\right)^2\,,
\end{equation}
with an effective per-observation uncertainty $\sigma_{{\rm eff},i} = (\sigma_{{\rm RV},i}^2 + \sigma_{\rm model}^2)^{1/2}$ that supplements the measurement noise $\sigma_{{\rm RV},i}$ with a term $\sigma_{\rm model} = 0.4\,$\kms\ accounting for residual imperfections of the PC reconstruction, set to the median RMSE of the $\mathcal{M}_{\rm RV}$ fits to the full VELOCE-I sample (see Section\,\ref{sec:rmse}). The prior on the PC coefficients is the conditional density $\mathcal{P}(p_1, \ldots, p_6 \mid \log P)$ calculated at the Cepheid's period, as described in Section\,\ref{subsec:prior_pcs}. No informative prior was placed on $v_\gamma$. Regarding $\Delta\phi$, for the 201 targets with a reliable photometric phase we adopted the conditional prior $\mathcal{P}(\Delta\phi \mid \log P)$ of Section\,\ref{subsec:prior_delta_phi}, while a uniform prior over  $\Delta\phi \in [-0.5, 0.5)$ was used otherwise. The resulting estimate was compared with the expected offset, $\Delta\phi_{E_G}$, between the photometric epoch and minimum radius calculated by replacing $E_{\rm Gaia\,DR3}$ with the applied $E_G$ in Eq.\,\ref{eq:real_phase_diff}.
Because the posterior is multimodal in $\Delta\phi$, a careful initialization of the optimization is essential. We sampled 300 candidate starting points over the prior domain of $(p_1, \ldots, p_6, \Delta\phi)$ using a seven-dimensional Latin Hypercube design \citep[\texttt{scipy.stats.qmc.LatinHypercube};][]{virtanen_scipy_2020}. For each candidate, we initialized $v_\gamma$ analytically as the weighted mean residual between the measurements and the pulsation curve prediction. The candidates were ranked by their log-posterior, and only the best $K$ were optimized in full, with $K$ decreasing from 30 for $N_{\rm RV} = 3$ to 5 for $N_{\rm RV} \geq 7$. The highest-posterior solution among them was retained as the MAP estimate for the subsample.

To summarize performance as a function of $N_{\rm RV}$, we characterized the central tendency and dispersion of $\Delta v_\gamma$, $\Delta P2P$, and the ratio $\Delta P2P/P2P_{\rm VELOCE}$ (labeled with the $N_{RV}$ subscript). For each $N_{\rm RV}$, the per-target distributions were combined into a single population-level distribution, from which both quantities were derived. To account for the different number of subsamples for V1197~Cen for $N_{\rm RV}= 3$ and 6, each target was assigned equal total weight by weighting every value in a target's distribution by the inverse of the number of subsamples available for that target. The weighted median of the combined distribution was adopted as the central tendency and its weighted MAD as the dispersion, both evaluated with a weighted-percentile estimator. The weighted median quantifies any systematic bias in the recovery, while the MAD provides a robust measure of the typical scatter expected for a Cepheid observed $N_{\rm RV}$ times.
We estimated the uncertainties on these summary statistics with a bootstrap over targets. In each of $10^4$ realizations, we resampled the set of targets with replacement, reconstructed the equal-weighted combined distribution from the resampled targets, and recomputed the weighted median and MAD. The dispersion of the bootstrap realizations was quantified as the half-width of the 16th to 84th percentile range, $HW = 0.5 \times (Q_{84} - Q_{16})$, and was adopted as the uncertainty on the corresponding statistic. Resampling at the level of targets, rather than of individual subsamples, is essential: the subsamples of a given target are mutually correlated and do not constitute independent draws, whereas distinct targets represent independent realizations of the underlying population.

We emphasize that this framework does not produce RV curve templates in the classical sense; rather, the mapping of $p_i$ across \logP\ enables an accurate fit to any number of RV measurements via MAP, and, as a limiting case, classical templates can nonetheless be recovered by adopting the maximum of each $\mathcal{P}(p_i \mid \log P)$ at a given \logP, albeit at some cost in accuracy.

\subsection{Evaluating the typical performance of the framework \label{subsec:result_framework}}

\begin{figure}[th!]
    \centering
    \includegraphics[width=.49\textwidth]{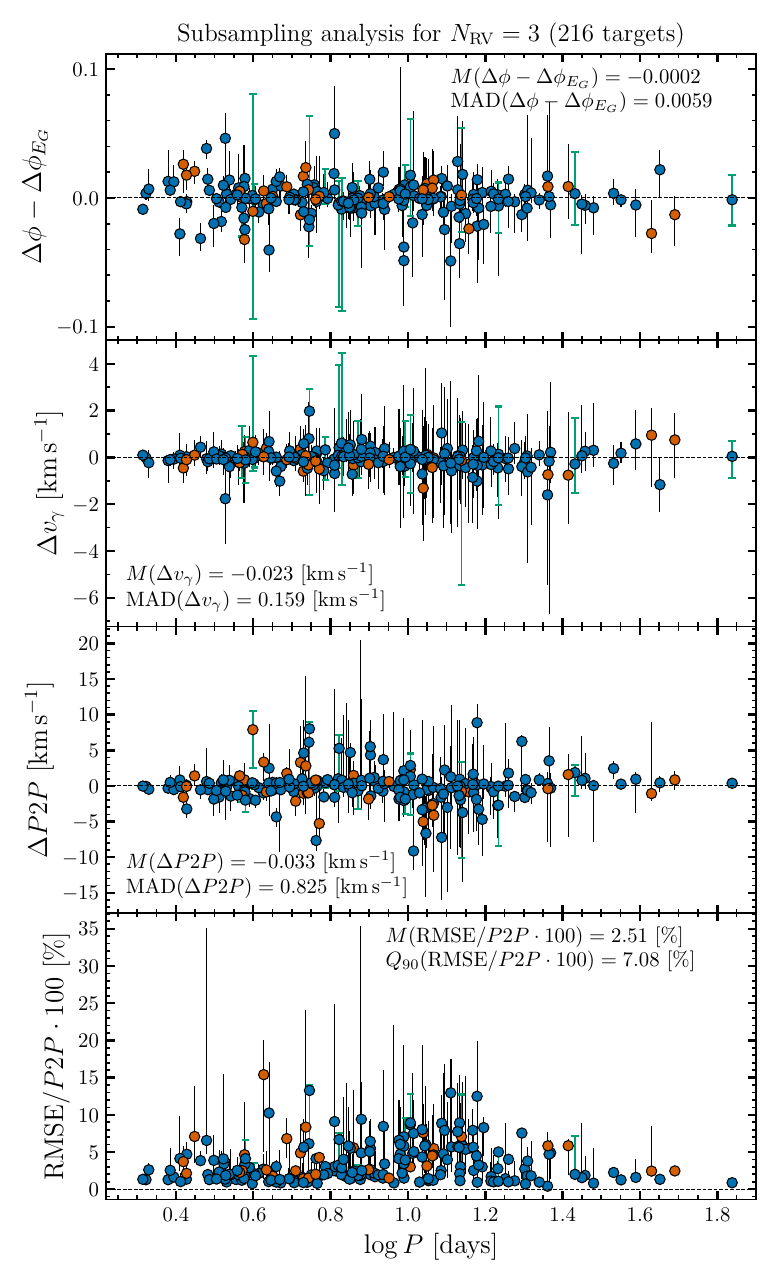}
    \caption{Performance of the PCs and priors applied to subsamples of size $N_{\rm RV}=3$. Markers show each target's median and the whiskers the range between the 16th and 84th percentiles of the results obtained across all subsamples; training and test targets are shown in blue and orange, respectively, as in the previous figures. Targets fitted without the informative prior $\mathcal{P}(\Delta\phi \mid \log P)$ (i.e. with a uniform prior on $\Delta\phi$) are marked by green whiskers. From top to bottom: the phase offset residual $\Delta\phi -\Delta\phi_{E_G}$, $\Delta v_\gamma$, $\Delta P2P$ and $RMSE/P2P_{VELOCE}$ (as a percentage). Each panel reports the median, $M$, and median absolute deviation, MAD, of the median's distribution, except the bottom panel, for which the median and 90th percentile, $Q_{90}$, are given. For clarity, $\alpha$\,UMi has been omitted from the bottom panel, where it is a strong outlier with $(RMSE/P2P \cdot 100) = 76.44\%$. Note that $M$ and MAD here refer to the distribution of per-target medians and therefore differ from the combined-distribution statistics of Table\,\ref{tab:nrv_summary}.\label{fig:subsampling_3}}
\end{figure}

Figures\,\ref{fig:subsampling_3}, \ref{fig:subsampling_examples_1}, and \ref{fig:subsampling_examples_2} present, for each target, the median and 16th-to-84th percentile ranges of several metrics obtained from $N_{\rm RV}$ observations, while Figure\,\ref{fig:example_fit_nrv} shows a few examples of the resulting RV curve reconstructions at $N_{\rm RV}=3$. Figure\,\ref{fig:subsampling_3} shows the results for the sparsest samples considered, $N_{\rm RV}=3$. Some iterations yielded poor fits owing to inadequate phase sampling. We intentionally retained these cases in order to maintain realism, since large surveys observe objects at essentially random phases. The reported percentile ranges therefore reflect such suboptimal cases and, for some targets, reveal markedly asymmetric and skewed distributions. 

The phase offset, $\Delta \phi$, and the pulsation average velocity, $v_\gamma$, are accurately recovered for all targets, even when only 3 observations are available. Both distributions reveal no meaningful bias, and the scatter is well contained within the uncertainties for both the training and test sets. The recovered $\Delta \phi$ is remarkably precise, with MAD$(\Delta \phi -\Delta \phi_{E_G}) \simeq 0.006$, i.e. better than 1\% of a pulsation cycle at $N_{RV}=3$. The percentile ranges are correspondingly narrow, such that the median of 149 of 216 (69\%) falls within its own 16th-to-84th percentile range of $\Delta \phi_{E_G}$. For the other targets the offsets, while formally exceeding this range, are negligible in absolute terms and have little to no appreciable consequence for the recovery of $v_\gamma$ and $P2P$.
The pulsation-averaged velocity is recovered in agreement with the reference $v_{\gamma,\rm VELOCE}$ for nearly all targets (211 out of 216, $\simeq 98\%$). As $N_{\rm RV}$ increases, the percentile ranges for $\Delta \phi$ narrow for all but a few targets, and those of $\Delta v_\gamma$ decrease steadily, as expected. 

\begin{table*}[t!]
\small
    \centering
    \caption{Recovery of RV curve properties from few observations \label{tab:nrv_summary}}
    \begin{tabular}{@{}ccccccc@{}}
    \toprule
       $N_{\rm RV}$ &   $N_{targets}$ & $M(\Delta v_{\gamma, N_{RV}})$ & $MAD(\Delta v_{\gamma, N_{RV}})$ & $M(\Delta P2P_{N_{RV}})$ & $MAD(\Delta P2P_{N_{RV}})$ & $Q_{90}(RMSE_{N_{RV}})$  \\
       &&  [\ms]   &  [\ms] &  [\ms]  &  [\ms]  & [\kms]  \\
    \midrule

          3 &         216 & $-21.6{\scriptstyle \pm12}$&  $496{\scriptstyle \pm 27} ~ (1.6\%)$ & $-88{\scriptstyle \pm69}$   &  $1178{\scriptstyle \pm 73} ~ (3.9\%)$  &  3.72 (10.4\%)  \\
          4 &         216 & $-3.6{\scriptstyle \pm8}$  &  $332{\scriptstyle \pm 19} ~ (1.1\%)$ & $-113{\scriptstyle \pm46}$  &  $874{\scriptstyle \pm 51} ~ (2.9\%)$   &  2.92 (7.8\%)  \\
          5 &         216 & $-4.3{\scriptstyle \pm7}$  &  $262{\scriptstyle \pm 15} ~ (0.9\%)$ & $-116{\scriptstyle \pm38}$  &  $724{\scriptstyle \pm 43} ~ (2.4\%)$   &  2.66 (6.7\%)  \\
          6 &         216 & $-4.2{\scriptstyle \pm5}$  &  $207{\scriptstyle \pm 12} ~ (0.7\%)$ & $-103{\scriptstyle \pm34}$  &  $615{\scriptstyle \pm 37} ~ (2.1\%)$   &  2.13 (5.6\%) \\
          7 &         215 & $-3.7{\scriptstyle \pm5}$  &  $192{\scriptstyle \pm 11} ~ (0.7\%)$ & $-97{\scriptstyle \pm28}$   &  $586{\scriptstyle \pm 37} ~ (2.0\%)$   &  2.32 (5.9\%)  \\
        10  &         213 & $-1.4{\scriptstyle \pm4}$  &  $137{\scriptstyle \pm 8} ~ (0.5\%)$  & $-77{\scriptstyle \pm27}$   &  $473{\scriptstyle \pm 30} ~ (1.6\%)$   &  1.69 (4.5\%)  \\
        15  &         205 & $-1.1{\scriptstyle \pm3}$  &  $92{\scriptstyle \pm 5} ~ (0.3\%)$   & $-62{\scriptstyle \pm26}$   &  $389{\scriptstyle \pm 30} ~ (1.3\%)$   &  1.30 (3.5\%) \\
        20  &         191 & $-3.0{\scriptstyle \pm2}$  &  $70{\scriptstyle \pm 4} ~ (0.2\%)$   & $-54{\scriptstyle \pm28}$   &  $380{\scriptstyle \pm 31} ~ (1.3\%)$   &  1.12 (3.0\%)  \\

    \midrule
    \bottomrule
    \end{tabular}
    \tablefoot{Summary of the application of the PCs and priors to subsamples of the available RV measurements. $N_{\rm RV}$ denotes the number of RV measurements per target, and $N_{targets}$ the total number of targets analyzed. $M(\Delta v_{\gamma, N_{RV}})$, $MAD(\Delta v_{\gamma, N_{RV}})$, $M(\Delta P2P_{N_{RV}})$ and $MAD(\Delta P2P_{N_{RV}})$ list the weighted median and $MAD$ values of the combined (across targets) distribution for $\Delta v_\gamma$ and $P2P$. Uncertainties were obtained via bootstrap resampling as discussed in Section\,\ref{subsec:methodology_map}. $Q_{90}(RMSE_{N_{RV}})$ reports the 90th-percentiles of the $RMSE$ calculated between the obtained fit and the \veloce\ original fit, $\texttt{V}^{FS}$. Values normalized by the $P2P$ and expressed in percentages are given in parentheses. }
\end{table*}

\begin{figure}
    \centering
    \includegraphics[width=0.48\textwidth]{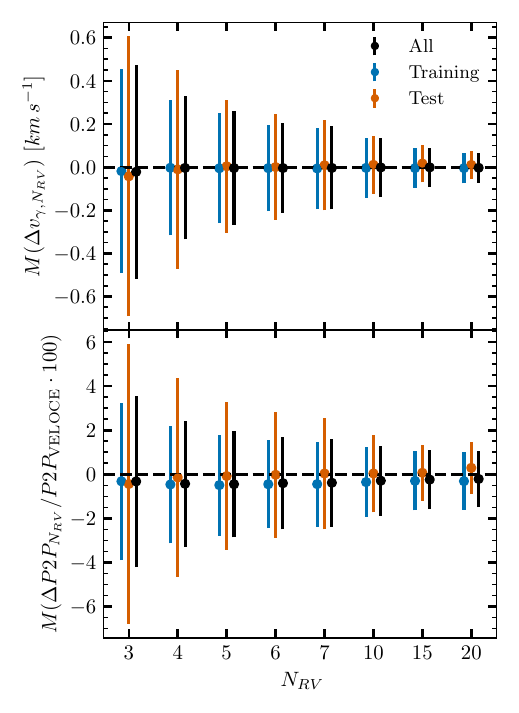}
    \caption{Accuracy of $v_\gamma$ (top) and $P2P$ (bottom) obtained via our framework fitting as a function of the number of available observations, $N_{\rm RV}$. Median and MAD values of the combined across-targets distribution are shown for the training (blue) and test (orange) sets, and their union (black).\label{fig:vgamma_p2p_vs_nrv}}
\end{figure}

\begin{figure}[th!]
    \centering
    \includegraphics[width=.49\textwidth]{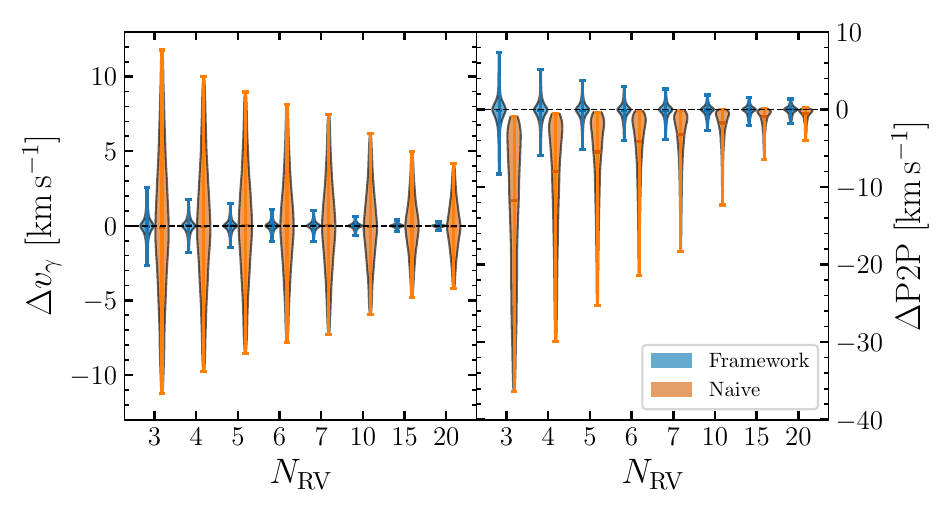}
    \caption{Comparison of the typical $\Delta v_\gamma$ and $\Delta P2P$ recovered via our fitting framework (blue) and via a naive estimator (orange), as a function of $N_{\rm RV}$. The naive estimates adopt the average and the maximum-minus-minimum of the RV measurements for $v_\gamma$ and $P2P$. The values shown correspond to the range between 5th and 95th quantiles.\label{fig:naive_comparison}}
\end{figure}

This performance is confirmed by the across-target central tendency and dispersion as a function of $N_{\rm RV}$ reported in Table\,\ref{tab:nrv_summary} and illustrated in the top panel of Figure\,\ref{fig:vgamma_p2p_vs_nrv}. The MAD$(\Delta v_{\gamma, N_{\rm RV}})$ decreases consistently with the number of available observations: from $\approx 500$\ms\ for $N_{RV}=3$ to roughly half that value ($\approx 260$\ms) for $N_{\rm RV}=5$ and to below $100$\ms for $N_{\rm RV}\ge15$. Figure\,\ref{fig:naive_comparison} compares these results with those obtained for the same subsamples using the 'naive' estimator of $v_\gamma$, namely the mean of the observations. Our framework clearly outperforms this benchmark at all $N_{\rm RV}$: the typical scatter is smaller by a factor of $\simeq 8$ at $N_{\rm RV}=3$, rising to $\simeq 20$ at $N_{\rm RV}=20$.

The recovery of $P2P$ likewise shows good overall performance with no globally apparent bias. As noted in Section\,\ref{subsubsec:p2p_all}, however, the reproduction of peak-to-peak amplitudes is hampered by the complexities inherent to the HP, notably for \logP$\in [1, 1.3]$. At low $N_{\rm RV}$ this manifests as an increased 16th-to-84th percentile range in that period, while for $N_{\rm RV}\geq 7$ the improved precision reveals the same systematic underestimation of $P2P$ found in the fits to the full dataset (Fig.\,\ref{fig:results_fit_rv_stats}; Section\,\ref{subsubsec:p2p_all}). This consistency confirms that the priors effectively constrain the fitted PC coefficients towards the values that would be obtained with complete information (i.e. all RV measurements), even for very sparse sampling.

The MAD$(\Delta P2P_{N_{\rm RV}})$ follows two distinct regimes as a function of $N_{\rm RV}$ (Table\,\ref{tab:nrv_summary}). It decreases rapidly from $\simeq 1.2$\kms\ ($\approx 4\%$) at $N_{\rm RV}=3$ to $\simeq 500$\ms\ ($\approx 1.6\%$) at $N_{\rm RV}=10$, and thereafter stabilizes near $\approx 400$\ms, a value that coincides with the median $RMSE$ of the fits to the full data set (Section\,\ref{sec:rmse}), and thus plausibly represents the accuracy floor set by the PC reconstruction itself. As shown in the bottom panel of Fig.\,\ref{fig:vgamma_p2p_vs_nrv}, the test set exhibits larger scatter for the sparsest samples, but converges to the training set values for $N_{\rm RV}> 7$. 
Comparison with the 'naive' estimator of $P2P$ (maximum-minus-minimum of RV measurements) reveals a clear improvement in both precision and agreement with $P2P_{\rm VELOCE}$ at low $N_{\rm RV}$. The advantage diminishes as $N_{\rm RV}$ grows, since additional epochs increase the likelihood of sampling phases near the true extrema: at $N_{\rm RV}=20$ the $P2P_{\rm VELOCE}$-normalized dispersion is comparable ($\simeq1.3\%$ for our framework against $\simeq1.5\%$ for the 'naive' estimator). 
Crucially, however, the naive estimator remains biased by construction, systematically underestimating $P2P$, while our estimates remain symmetric about $P2P_{\rm VELOCE}$ (Fig.\,\ref{fig:naive_comparison}). For low-amplitude targets ($P2P< 10$\kms), the advantage of the framework over the naive estimator is reduced: the amplitude-driven scatter of the naive estimates is intrinsically small for these stars, while the PC prior tends to pull the recovered $P2P$ towards the typical amplitude at that period. Even in this regime, however, the naive $P2P$ estimator remains biased low by construction.

The overall agreement between $\mathcal{M}_{\rm RV}$ and $\texttt{V}^{FS}$, as measured by the $RMSE$, depends on both $N_{\rm RV}$ and $\log P$, likely as a consequence of the correlated trend with $P2P$. RV curves for $\log P< 0.6$ are reproduced to within $< 1$\kms\ or better even at $N_{RV}=3$, the sole exception being $\alpha$\,UMi ($RMSE\simeq 2.7$\kms\, or $\approx 76.4\%$ of $P2P_{\rm VELOCE}$). Globally, the $RMSE$ within which 90\% of the Cepheids are reproduced improves from $\simeq 3.7$\kms\ at $N_{\rm RV}=3$ to $\simeq 2.3$\kms\ at $N_{\rm RV}=7$ and $\simeq 1.1$\kms\ at $N_{\rm RV}=20$. Analogously, the corresponding 90th percentile of $RMSE/P2P_{VELOCE}$ improves from 10.4\% to 5.9\% and to 3\%, respectively.
The $RMSE/P2P_{VELOCE}$ distribution reports median values below $ 15\%$ at $N_{\rm RV}=3$ for every target except $\alpha$\,UMi, but the percentile ranges are particularly large and skewed for a few targets: $\alpha$\,UMi, IT\,Car, ASAS\,J064540+0330.4, V0378\,Cen, X\,Lac and HW\,Car all have the 84th percentile above 20\%. All these targets exhibit low P2P, to varying degrees, relative to most Cepheids with similar $\log P$. For such stars, the PC-prior favors solutions that tend towards the typical RV curve at that period and therefore towards larger $P2P$. The effect is more pronounced at low $N_{\rm RV}$, where the prior dominates, and diminishes as observations accumulate, with the exception of $\alpha$\,UMi, which remains an outlier at all $N_{\rm RV}$.

Three targets, EU\,Tau, V0371\,Gem and DK\,Vel, exhibit counter-intuitive behavior: their $RMSE/P2P_{VELOCE}$ and $\Delta \phi$ are well recovered for the sparsest subsamples, whereas for $N_{\rm RV}\gtrsim7$ the scatter of the results increases substantially. Inspection of the corresponding fits reveals spurious oscillations ("ringing") in the reconstructed RV curves between the sampled phases, which are absent at lower $N_{\rm RV}$. This behavior reflects the changing balance between likelihood and prior in the MAP objective: the data term scales with $N_{\rm RV}$, whereas the prior term does not. For few observations the fit is prior-dominated and the higher-order coefficients remain confined near their most probable values, yielding smooth, template-like reconstructions. As $N_{\rm RV}$ increases, the likelihood progressively dominates and the higher-order PCs are free to adjust to the individual measurements. Since the phase coverage remains incomplete, this produces oscillations between the observed epochs which, although they do not degrade the residuals at the sampled phases, increase the $RMSE$ computed against $\texttt{V}^{\rm FS}$ over the full phase grid and render the $\Delta \phi$ less constrained (see Fig. \,\ref{fig:subsampling_examples_2}). We note that this limitation is not intrinsic to the framework and could likely be mitigated by regularizing the higher-order PC coefficients, by tempering the prior, or by refining the optimization strategy.

In summary, our framework recovers $v_\gamma$ from merely three observations with a typical scatter of $\simeq 0.5$\,\kms\ and a bias ($\leq 22$\ms) comparable to the per-epoch precision of \veloceDRI\ measurements. Applications demanding better than
$0.2\,$\kms\ (0.1\,\kms) require $N_{\rm RV} \gtrsim 7$ (15). The peak-to-peak amplitude is less well constrained, as expected, yet is already recovered to within 4\%
of its value from three observations, improving to below 2\% for $N_{\rm RV} > 7$. The reconstruction of the RV curve as a whole improves substantially with the number of epochs: the RMSE within which 90\% of Cepheids are reproduced decreases from $\simeq 3.7$\,\kms\ (or 10.4\% of the total amplitude) at $N_{\rm RV}=3$ to $\simeq 2.3$\,\kms\ ($\simeq\,6\%$) at $N_{\rm RV} = 7$. These figures serve a dual purpose. For large surveys, in which the number of epochs per target is fixed by scheduling constraints, they allow the performance of our framework to be anticipated for a given cadence. Conversely, for smaller, dedicated programs, in which the number of epochs remains at the observer's discretion and telescope time is a scarce resource, they provide a direct basis for defining the cadence required to reach a target precision on $v_\gamma$ or $P2P$.

\section{Application to extragalactic Cepheids}\label{sec:extragal}
\label{subsec:extragalactic}

We applied our framework to time-series RV measurements of Cepheids in the LMC and SMC (cf. Section\,\ref{subsec:extragal_data}) in order to test its performance at a metallicity systematically lower than the \veloceDRI\ data set, which is mostly comprised of near Solar metallicity Cepheids (cf. Fig.\,20 in \veloceDRI, which implies an average near $+0.1$ with considerable scatter), whereas the mean metallicity of SMC and LMC Cepheids is lower, $\mathrm{[Fe/H]_{LMC}} = -0.409 \pm 0.076$ \citep{romaniello_iron_2022} and $\mathrm{[Fe/H]_{SMC}} = -0.785\pm 0.082$ \citep[citing Romaniello et al. in prep.]{breuval_small_2024}. Additionally, we modified our methodology to specifically search for spectroscopic binaries (Section\,\ref{subsec:method_MC}), and sought to obtain a first view of the line-of-sight velocity distributions of both Clouds. The results for the LMC and SMC are presented in Sects.\,\ref{subsec:LMC} and \ref{subsec:SMC}, respectively.

\subsection{Literature RVs of extragalactic Cepheids}\label{subsec:extragal_data}

\cite{storm_calibrating_2011} published 509 precise RV measurements (per-epoch uncertainty $\le 100$\,\ms) of 22 fundamental mode Cepheids in the Large Magellanic Cloud (LMC) based on observations with HARPS \citep{2003Msngr.114...20M} and FEROS \citep[used only for HV914 and HV12717;][]{1999Msngr..95....8K}. Most spectra were collected during a single 16-night observing run,  with additional epochs obtained over the following two seasons to fill phase gaps. While this strategy yielded densely sampled RV curves, the relatively short temporal baseline limits the detection of possible orbital motion. Indeed, \cite{storm_calibrating_2011} identified evidence of orbital motion only for HV\,1005, where two observations taken two years after the main campaign appear to be shifted from the rest of the RV curve.

As a second dataset, we adopted the RV measurements published by \cite{gieren_effect_2018} for Cepheids in the Small Magellanic Cloud (SMC). This sample comprises 714 RV measurements of 26 Cepheids obtained using 3 high resolution spectrographs: HARPS, MIKE \citep{2003SPIE.4841.1694B} and UVES \citep{2000SPIE.4008..534D}. These RV measurements have typical per-epoch uncertainties of $\sim250$\ms and were collected between 2012 and 2017, providing observational baselines sufficient to detect long-term orbital trends. \cite{gieren_effect_2018} included 5 additional targets from \cite{storm_bvrijk_2004} and HV\,837 from \cite{imbert_radial_1989}. We did not consider these stars here, restricting our analysis to the newly acquired data.

\cite{gieren_effect_2018} reported velocity variations indicative of orbital motion in six of the 26 Cepheids: OGLE\,SMC-CEP-1680, -1686, -1693, -1729, 1977, and  -2905. 
Among these, SMC-CEP-1686 and SMC-CEP-1693 exhibited particularly complex behavior that prevented the identification of a consistent systemic velocity. In the case of SMC-CEP-1686, concurring with the authors, we noted that the RV variations cannot be described by simple additive offsets, whereas for SMC-CEP-1693, the variations are so large relative to the pulsational amplitude that no sensible RV curve can be identified. While \cite{gieren_effect_2018} analyzed SMC-CEP-1686 without adopting any correction and excluded SMC-CEP-1693 from their analysis, we opted to remove both stars. For the remaining 4 potential binaries, \cite{gieren_effect_2018} minimized the effect of orbital motion by applying epoch-dependent velocity offsets to corresponding data.

\subsection{Time-resolved hierarchical fitting and detection of orbital motion}\label{subsec:method_MC}
To identify spectroscopic binaries (SB) among the two datasets, we partitioned each RV time series into temporal clusters following  the approach of \citet[][and Section\,5.1 in \veloceDRI]{Anderson2016rvs}. This procedure enables the determination of the pulsation-averaged RV for each cluster, and allows the assessment of its temporal variability, thus retaining sensitivity to orbital motion on longer timescales. As done in \veloceDRI, clustering was performed using KDE with the bandwidth parameter defined as $\max(80, \min(3\cdot P, 100))$d. Each case was visually inspected, and the bandwidth and/or number of clusters were adjusted to ensure adequate grouping and a minimum of 2 RV measurements per cluster.

We fitted 2 models to the RV measurements. The first, $\mathcal{M}_{\rm RV}^{\rm SOLO}$, assumes no orbital motion and fits only the pulsation component as defined by Eq.\,\ref{eq:map_model}. The second, $\mathcal{M}_{\rm RV}^{\rm MULTI}$, accounts for possible orbital motion, by fitting a hierarchical model that includes both the pulsational component and cluster-specific pulsation average velocity, $v_{\gamma,\mathrm{cluster}}$. This formulation enables the quantification of variations in the mean velocity over time. Importantly, the pulsation model is shared across all clusters, ensuring that the full dataset contributes to constraining the PC coefficients. For both models, as explained in Sections\,\ref{subsec:prior_delta_phi} and \ref{subsec:methodology_map}, we fitted $\Delta \phi$, to ensure $\phi=0$ occurs at minimum radius.
The pulsation component was constrained using the previously derived priors (see Section\,\ref{subsec:prior_pcs}). However, given the high dimensionality and potential parameter degeneracies, we performed the optimization in two steps. First, a global search using the Differential Evolution (DE) algorithm located the best-fit region of the parameter space. Second, a Markov Chain Monte Carlo (MCMC) exploration refined the parameter estimates and characterized their posterior distributions. This strategy provided robust constraints on both the pulsation curve and $v_{\gamma, \mathrm{cluster}}$. 

A Cepheid can be identified as a binary when the scatter among its own $v_{\gamma,\mathrm{cluster}}$ values significantly exceeds the precision with which a single $v_{\gamma,\mathrm{cluster}}$ can be determined, $\sigma_{\rm cluster}$. This precision is not known a priori, as it depends both on the quality of the RV measurements and on the ability of our framework to recover $v_\gamma$ from the epochs available in each cluster. It cannot simply be estimated from the observed dispersion of the sample either, since that dispersion is itself inflated by the binaries we seek to identify. To break this circularity, we estimated $\sigma_{\rm cluster}$ from the dispersion of the sample after excluding the E[$N_{\rm{SB1}}$] targets exhibiting the largest scatter, where E[$N_{\rm{SB1}}$] is the expected number of SB1 systems in the sample.

We determined E[$N_{\rm{SB1}}$] from the properties of the orbital solutions determined in \veloce-II and the temporal sampling of the LMC and SMC datasets, taking into account also the precision of the fitting methodology. We adopted the average $RMSE$ of the $\mathcal{M}_{\rm RV}^{\rm MULTI}$ fits, $\overline{RMSE}(\mathcal{M}_{\rm RV}^{\rm MULTI})$, as a representative measure of the overall precision of the PC-based models. First, we sampled each \veloce-II orbit at the mean BJD of each cluster for a randomized orbital phase and added Gaussian noise with a standard deviation of $ \overline{RMSE}(\mathcal{M}_{\rm RV}^{\rm MULTI})$, scaled by the square root of the number of observations per cluster. Orbital motion was deemed ``detected'' if the simulated cluster-to-cluster $P2P$ variation exceeded twice the $\overline{RMSE}(\mathcal{M}_{\rm RV}^{\rm MULTI})$. We repeated this process 1000 times to estimate the detection efficiency for each \veloce-II orbit. The mean of the detection efficiency across all \veloce-II orbital solutions was adopted as the representative detection efficiency for the temporal configuration of the dataset. To quantify the false-alarm probability (FAP) due to noise alone, we repeated the procedure using a constant (non-orbital) RV function.
Second, we considered the RV trends reported in Tab. 5 of \veloce-II and qualitatively estimated which of them could be recovered given the dataset average observational baseline and $\overline{RMSE}(\mathcal{M}_{\rm RV}^{\rm MULTI})$.
Lastly, we considered SB1s in \veloce-II detected solely via the RV Template Fitting (RVTF) unidentifiable within the LMC and SMC datasets because of their rather short temporal baselines.
The typical precision, $\sigma_{\mathrm{cluster}}$, was thus computed based on the sample of Cepheids reduced by the $E[N_{\rm{SB1}}]$ exhibiting the largest variations in $v_{\gamma,\mathrm{cluster}}$: 

\begin{equation}\label{eq:sigma_intr}
    \sigma_{\mathrm{cluster}} = \sqrt{\frac{\sum_{\rm i}\sum_{\rm cluster} (v_{\gamma, \rm cluster, i} -\overline{v_{\gamma, \rm cluster, i}})^2}{N_{\rm clusters, No\ SB1} - N_{\rm cepheids, No\ SB1}}}
\end{equation}
where $N_{\rm cepheids, No\ SB1} = N_{\mathrm{cepheids}} - E[N_{\rm{SB1}}]$, $N_{\rm clusters, No\ SB1}$ is the  total number of temporal clusters for the $N_{\rm cepheids, No\ SB1}$ targets, and $\overline{v_{\gamma, \rm cluster, i}}$ the mean velocity of all clusters for the $i$th Cepheid. 

Finally, we considered as SB1 candidates any Cepheids with at least one $3\sigma_{\rm cluster}$ outlier cluster measurement, i.e., where:
\begin{equation}
    f_{\sigma, \rm cluster} = \max(|v_{\gamma, \rm cluster} -\overline{v_{\gamma, \rm cluster}}|) / \sigma_{\rm cluster} \geq 3 \ .
\end{equation}
The following reports the results from the $\mathcal{M}_{\rm RV}^{\rm SOLO}$ model for stars where no orbital motion was detected; for candidate SB1s, the results from the $\mathcal{M}_{\rm RV}^{\rm MULTI}$ analysis are presented.

\subsection{LMC}\label{subsec:LMC}
Since most of the observations from \citet{storm_calibrating_2011} were collected during a single observing run, only 12 of the 22 LMC Cepheids yielded more than one temporal cluster. Furthermore, the limited baseline prevented an adequate application of our methodology to estimate $E[N_{\rm{SB1}}]$ and thus $\sigma_{\rm cluster}$ for this dataset.

We fitted the $\mathcal{M}_{\rm RV}^{\rm SOLO}$ model to all targets and the $\mathcal{M}_{\rm RV}^{\rm MULTI}$ model to the 12 targets with multiple clusters. 
The fit for HV877 indicated that the measurement at $\rm{BJD}\simeq54046.71$ is a potential outlier, which we excluded from the analysis.

Figure\,\ref{fig:kde_components_MC} shows the fitted PC coefficients for each LMC Cepheid in yellow; their distribution closely follows the one of the training set used in Section\,\ref{subsec:prior_pcs} without any significant outliers. Visual inspection of the fitted RV curves revealed overall good agreement, with median $RMSE=0.79$\kms for the simpler $\mathcal{M}_{\rm RV}^{\rm SOLO}$ model.

\begin{figure}
    \centering
    \includegraphics[width=0.42\textwidth]{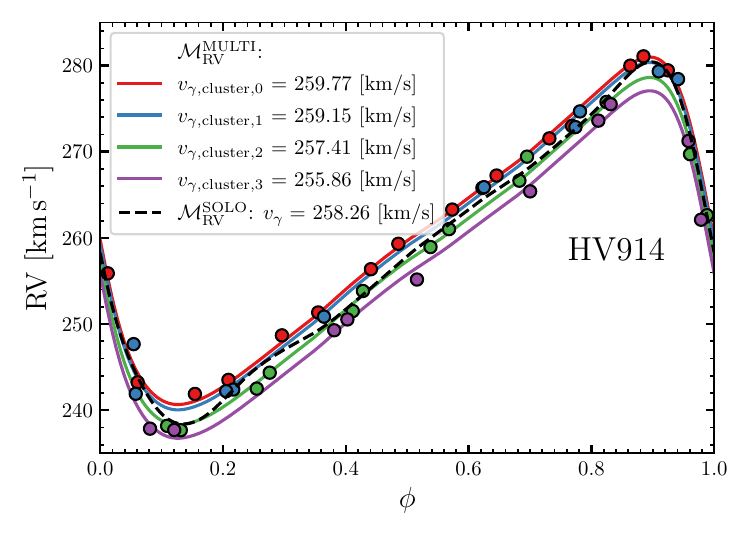}
    \includegraphics[width=0.42\textwidth]{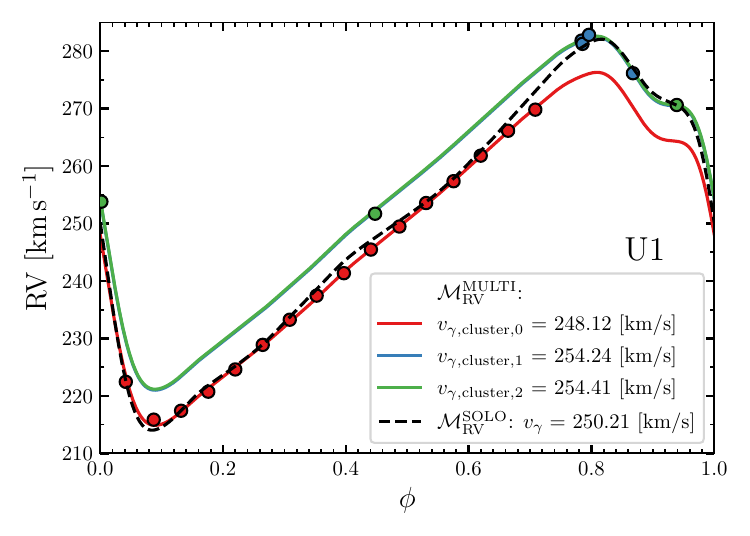}
    \includegraphics[width=0.42\textwidth]{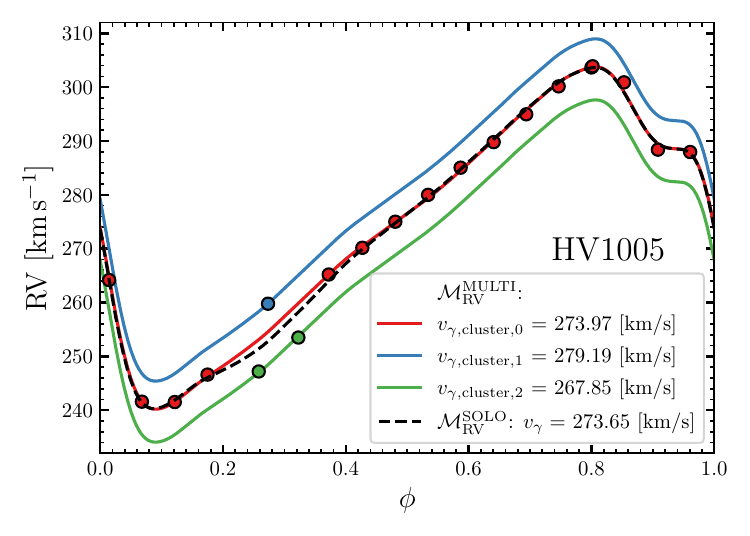}
    \caption{RV measurements and model fits for HV\,914 (top), U1 (middle) and HV1005 (bottom). The data points are color-coded by temporal cluster, with corresponding $\mathcal{M}_{\rm RV}^{\rm MULTI}$ cluster models shown in matching colors. The overall single-epoch fit, $\mathcal{M}_{\rm RV}^{\rm SOLO}$, is displayed for comparison.\label{fig:HV914_and_U1}}
\end{figure}

\begin{figure}
    \centering
    \includegraphics[width=0.5\textwidth]{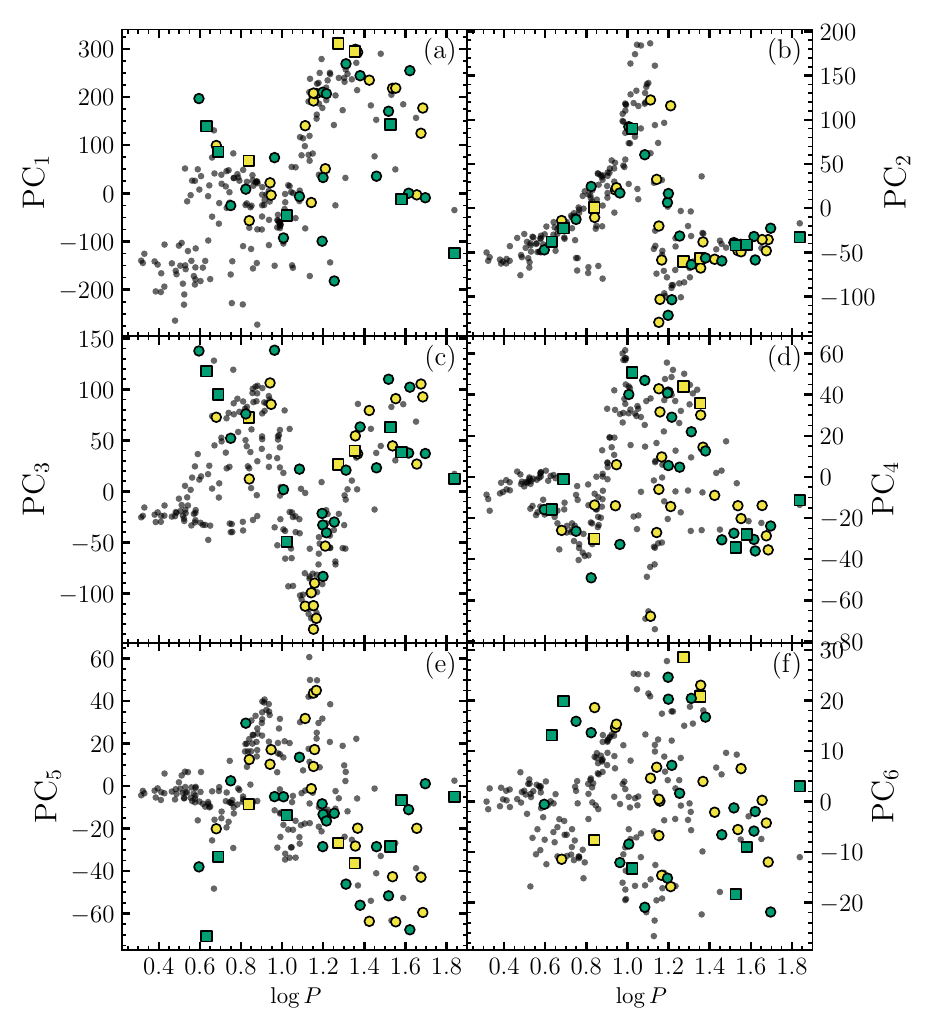}
    \caption{\label{fig:kde_components_MC} Distribution of the fitted coefficients for the LMC (in yellow) and SMC (in green) targets plotted over the training set used to determine the prior via KDE in Section\,\ref{subsec:prior_pcs}. Determined and suspected binaries are presented as filled squares, whereas the rest are plotted as filled circles.}
\end{figure}

The $\mathcal{M}_{\rm RV}^{\rm MULTI}$ model yielded substantially better results than the $\mathcal{M}_{\rm RV}^{\rm SOLO}$ model only for HV914, HV1005, and U1 (OGLE\,LMC-CEP-79); their $RMSE$ decreasing from 1.84 to 1.06\kms\ for HV914, from 2.85 to 0.83\kms\ for HV1005, and from 1.91 to 0.59 \kms\ for U1. 
As the panels in Fig.\,\ref{fig:HV914_and_U1} illustrate, the detection of orbital motion appears robust for HV914 and U1. 
For HV1005, we confirmed the findings of \cite{storm_calibrating_2011}, as two observations collected years after the main campaign are vertically offset from the rest. While this may be indicative of orbital motion, more observations are required to exclude other potential causes, such as phase shifts, which O-C analyses of photometric data can efficiently identify \citep{Rathour2024}.

Adopting the $\mathcal{M}_{\rm RV}^{\rm MULTI}$ fit for these 3 targets and the  $\mathcal{M}_{\rm RV}^{\rm SOLO}$  for all others, we found median and 90th percentile of the $RMSE$ to be 0.78\kms and 1.16\kms, respectively, fully consistent with the expected performance and capabilities of the PCs (see Fig.\,\ref{fig:results_fit_rv_stats}e and Tab.\,\ref{tab:nrv_summary} for $N_{\rm RV}\ge 15$). The complete set of results is provided in  Tab.\,\ref{tab:extragalactic}.

\begin{figure*}
    \centering
    \includegraphics[width=0.8\textwidth]{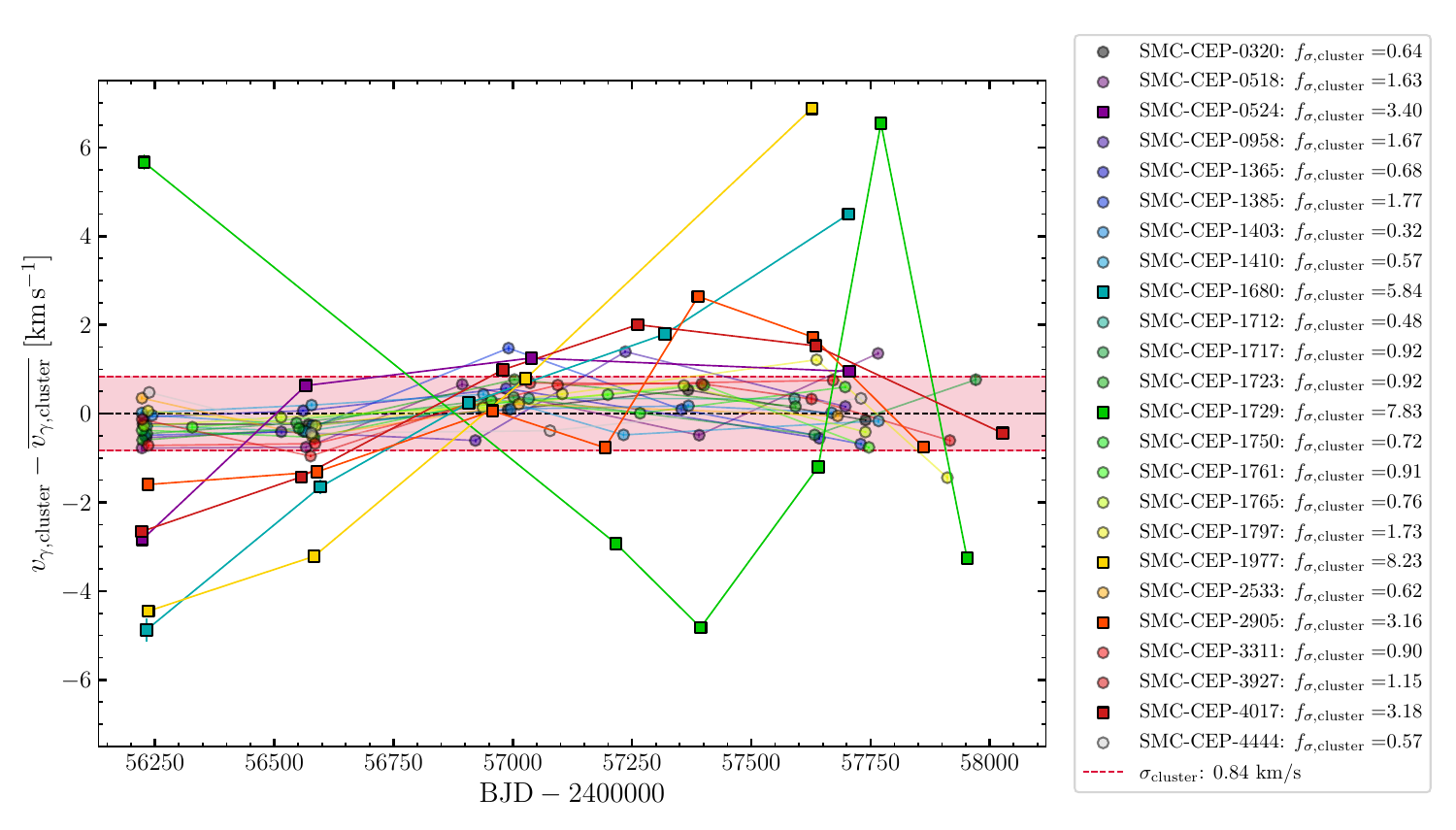}
    \caption{\label{fig:offsets_vs_mean} Per target residuals of cluster-specific pulsation average velocities, $v_{\gamma,\rm cluster} - \overline{v_{\gamma,\rm cluster}}$, as a function of time for the SMC sample. The red shaded region indicate $\sigma_{\rm cluster}$, while $f_{\sigma, \rm cluster}$ values for each target are listed in the legend. Targets with $f_{\sigma, \rm cluster}> 3$, the adopted threshold for identifying SB1 candidates, are marked with squares; all other targets are shown as circles.}
\end{figure*}

\subsection{SMC}\label{subsec:SMC}
Sufficiently long baselines to allow clustering of the observations were available for all 24 Cepheids under analysis, $N_{\rm cepheids}^{\rm SMC}$. We fitted both $\mathcal{M}_{\rm RV}^{\rm SOLO}$ and $\mathcal{M}_{\rm RV}^{\rm MULTI}$ models to all objects, obtaining a $\overline{RMSE}(\mathcal{M}_{\rm RV}^{\rm MULTI}) = 1.21$\kms. The results are tabulated in Tab.\,\ref{tab:extragalactic} and the fitted coefficients are shown in green in Fig.\,\ref{fig:kde_components_MC}. As for the LMC's Cepheids, the coefficient distributions globally follow the one of the \veloce\ targets, albeit with some differences at the longest and shortest periods, especially for PC$_1$ and PC$_5$. The outlier in $\rm PC_{5}$ corresponds to SMC-CEP-1729 ($\log P \sim 0.6$); however, visual inspection confirmed that the obtained fit provides a reasonable reproduction of the RV curve.

We simulated the detectability of the \veloce-II\ orbital solutions under the temporal sampling of the \citet{gieren_effect_2018} data set and estimated the associated false alarm rate, finding that $\sim 90.6\%$ of the 33 \veloce-II\ orbits would be recovered with a FAP $\sim 4.0\%$, yielding a net detection rate of $\sim 86.6\%$. We then considered the long-term RV trends reported in \veloce-II\ and concluded that between 9 and 12 of these could likewise be detected. Combining the two, the expected number of SB1 systems in the \citet{gieren_effect_2018} sample is
\begin{equation}
    \label{eq:expected_n_binaries}
    E[N_{\rm{SB1}}] = \frac{0.866\cdot33 + N_{\rm trend}}{N_{\rm cepheids}^{\rm VELOCE}} \; N_{\rm cepheids}^{\rm SMC} \simeq 3.8\;(3.5)
\end{equation}
where ${N_{\rm cepheids}^{\rm VELOCE}} = 258$ is the number of Galactic classical Cepheids in \veloceDRI\ and $N_{\rm trend} = 12\;(9)$ denotes the optimistic (conservative) number of recoverable long-term trends, and the corresponding values of $E[N_{\rm SB1}]$ are given in the same order. We therefore adopt $E[N_{\rm SB1}] = 4$.

Figure\,\ref{fig:offsets_vs_mean} illustrates $v_{\gamma,\mathrm{cluster}}$ as a function of time for the SMC sample. The 4 targets exhibiting the largest $v_{\gamma, \mathrm{cluster}}$ scatter are: SMC-CEP-1680, -1729,-1977, and -4017. 
We computed the scatter of the cluster-averaged velocities, $v_{\gamma, \mathrm{cluster}}-\overline{v_{\gamma, \mathrm{cluster}}}$, excluding these 4 targets from the calculation and applying Eq.\,\ref{eq:sigma_intr}, obtaining $\sigma_{\rm cluster} = 0.84$\kms.

In total, 6 Cepheids exhibit $v_{\gamma,\mathrm{cluster}}$ scatter exceeding $3\sigma_{\rm cluster}$: SMC-CEP-1680, -1729, -1977, -4017, -2905, and -0524. \citet{gieren_effect_2018} previously identified SMC-CEP-1680, -1729, -1977, and -2905, as SB1 Cepheids, and our methodology confirms their result. Following visual inspection of the $\mathcal{M}_{\rm RV}^{\rm MULTI}$, we concluded that SMC-CEP-4017 does show evidence of orbital motion. In contrast, although SMC-CEP-0524 formally meets the $f_{\sigma, \rm cluster}$ threshold, its RV measurements seem to follow a smooth and plausible RV curve at that pulsational period that the model does not accurately reproduce. Its SB1 nature hinges on the first temporal cluster, whose RV measurements sample phases not covered by the other clusters, preventing a direct comparison between epochs. After careful evaluation, we concluded that the star is probably not a binary but further observations would nevertheless be required to fully rule out binarity.
In summary, we identified five binary Cepheids within the sample, including one new detection (SMC-CEP-4017), consistent with the expected number $E[N_{\rm{SB1}}]$ ($4\pm 2$; Poisson error), previously determined.

The fit accuracy was found to be overall lower than for the LMC, with median and 90th percentile $RMSE$ values of 1.15\kms and 1.89\kms, respectively. The increase in $RMSE$ may be attributed to a combination of factors, including unaccounted instrumental offsets and/or discrepancy, unidentified binaries, period changes, and the larger metallicity difference relative to the \veloce\ RV curve from which the priors were derived.

\subsection{LMC and SMC kinematics traced by Cepheids\label{sec:kinematics}}

\begin{figure*}
    \centering
    \includegraphics[width=0.495\textwidth]{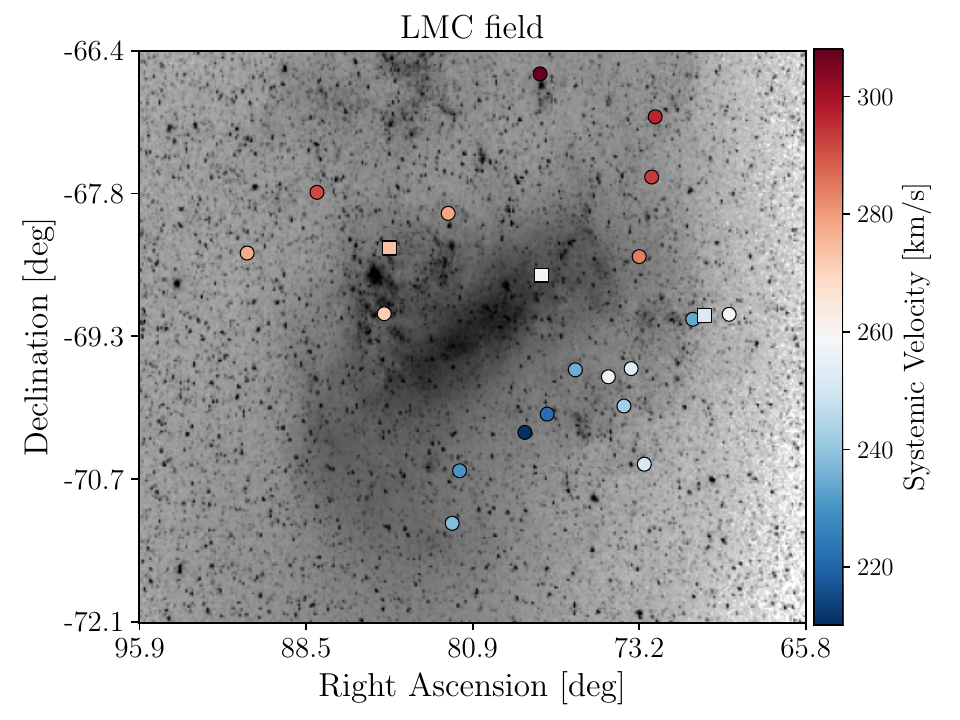}
    \includegraphics[width=0.495\textwidth]{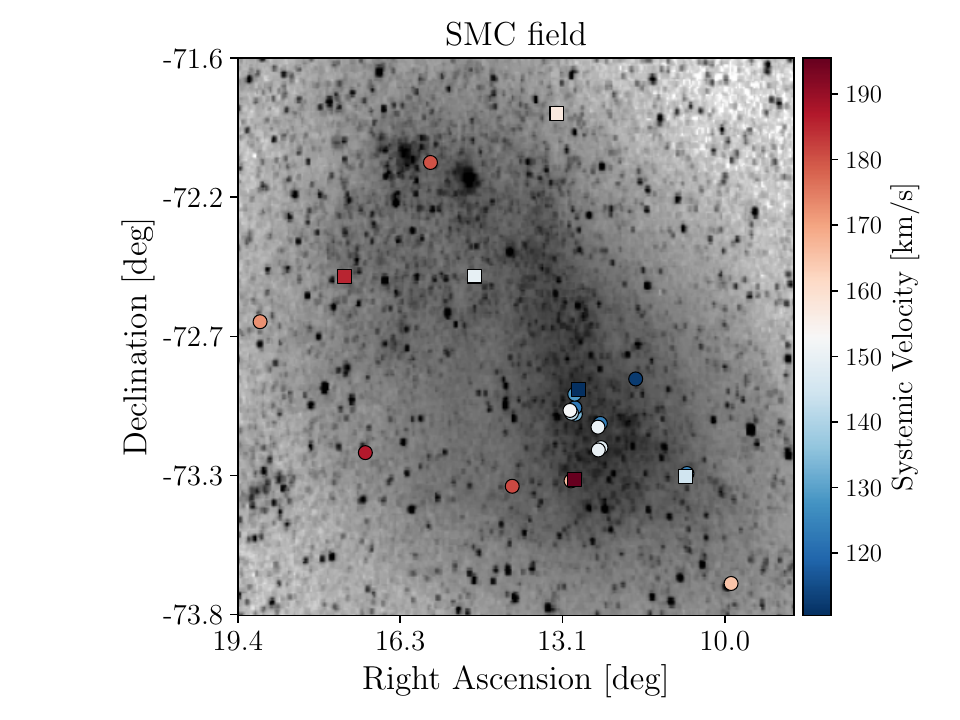}
    \caption{\label{fig:lmc_map}Line of sight velocities of Cepheids across the LMC (left) and the SMC (right) as determined using observations from \citet{storm_calibrating_2011} and \citet{gieren_effect_2018}. Spectroscopic binaries are shown using squares, other targets as  circles. The line-of-sight systemic velocity is shown using a color scale. The background image was created using the Mellinger survey \citep{mellinger_color_2009} (\textit{Mellinger-Green}) available from the \texttt{astroquery.skyview python} package \citep{ginsburg_astroquery_2019}.}
\end{figure*}

Figure\,\ref{fig:lmc_map} displays the spatial distribution of the analyzed Cepheids, color-coded by their systemic velocities in both galaxies tabulated in Tab.\,\ref{tab:extragalactic}. While the samples are small compared to targeted studies of LMC and SMC dynamics, some interesting features are evident. 
In the LMC, we find an interesting line-of-sight pattern across the field that resembles the rotation pattern reported by detailed analyses from the literature \citep{vanderMarel2014,gaiadr3.radvel,gaiadr3.lmckinematics}. 
In the SMC, the spatial distribution of the Cepheids does not clearly indicate a large-scale rotation. We note that several targets around $RA\sim 13.1$[deg] may trace a localized rotational pattern within the SMC's core, where line-of-sight effects are substantially reduced \citep{breuval_small_2024}. However, firmly establishing any rotational signatures within the SMC requires larger samples due to the SMC's structure \citep[cf.][]{Scowcroft2016, Jacyszyn-Dobrzeniecka2016} and the dynamical interactions with the LMC \citep[e.g.,][]{DeLeo2020}. Further understanding the SMC's kinematics is thus of great interest for understanding the Magellanic system \citep[cf.][and references therein]{Zivick2021}. 
4MOST will soon enable a detailed mapping of the line-of-sight velocity of many thousands of Cepheids that will be useful for understanding the dynamical histories of both galaxies. The usefulness of Cepheids, specifically, in such endeavors derives from their relatively young ages (roughly $30 - 600$\,Myr), which can be determined precisely on a star-by-star basis using period-age relations \citep{Anderson2016rot}, their high luminosity, which enables good RV precision, well-established LMC/SMC membership based on the Cepheid period-luminosity relation, and even the ability to constrain the position along the line of sight within each galaxy, especially within the elongated structure of the SMC \citep[e.g.,][]{Scowcroft2016}. Given the high accuracy achievable for individual Cepheids, we consider our methodology very promising to pursue highly detailed, population-specific kinematic studies of the Magellanic Clouds using 4MOST, which will begin survey operations in 2026.

\section{Conclusions}\label{sec:conclusion}

We have developed a framework for accurately fitting Cepheid RV curves, based on a PCA of the high-precision data set from \veloceDRI. Our framework differs from a typical template fitting approach in that prior distributions of the RV curve coefficients are used to fit RV measurements instead of adopting fixed RV curve parameters for a given \logP. Classical templates can nonetheless be derived from our framework, e.g., by adopting the maximum of each prior distribution for a given \logP, albeit at the cost of accuracy. The PCs, $\overline{\texttt{V}^{\rm FS}}$, the PC coefficients and the prior distributions derived from the PC coefficients and from $\Delta \phi$ are published in electronic form alongside this article. Furthermore, the complete code used for this analysis, together with \texttt{Python} routines for applying the framework, is currently available on \texttt{GitHub}\footnote{\href{https://github.com/giordiviv/veloce-framework}{\nolinkurl{github.com/giordiviv/veloce-framework}}} and will be archived on the Zenodo VELOCE repository\footnote{\href{https://zenodo.org/records/12818503}{\nolinkurl{zenodo.org/records/12818503}}} upon publication.

We extensively tested our methodology and found overall excellent performance, indicating that the PCs capture the diversity of Cepheid RV curves in scale and morphology very well. Our method allows us to determine $v_\gamma$ (pulsation average velocities) to within a couple tens of \ms, slightly better than the per-epoch precision of the RV measurements used, and with no significant bias (median offset of $5\,$\ms). Peak-to-peak amplitudes are usually recovered to within  $\sim 1.1\%$. Limitations arising near the complex HP and in the case of peculiar Cepheids with unusually low amplitudes were discussed. The diversity of RV curve shapes in the interval \logP$\in [1,1.3]$ prevents a unique mapping of RV curve shape based on period and results in larger differences in $P2P$ than at other periods, systematically biased low by a few per cent (Fig. \ref{fig:results_fit_rv_stats}d). Cepheids exhibiting unusually low $P2P$ show higher-than-usual RMSE.

By subsampling the \veloce\ data set, we demonstrated that $v_\gamma$ and $P2P$ are typically recovered within 0.5\kms\ and 4\%, respectively, from as few as 3 observations. The phase offset $\Delta\phi$ is recovered without significant bias at all $N_{\rm RV}$, with any residual offsets remaining negligible in absolute terms, and we note that the prior on the phase further constrains the per-target scatter. This highlights the practical usefulness of our fitting framework in the context of large spectroscopic surveys, such as 4MOST or WST.

Analyzing literature time-series RV measurements of Cepheids in the Magellanic Clouds, we obtained accurate RV curve fits with 90th-percentile $RMSE$ values of $1.16$\kms\ for the LMC and $1.89$\kms\ for the SMC. We carefully determined the ability of our method and the available data to identify spectroscopic binaries. Using a hierarchical model, we identified several previously reported binary systems and three new detections (two in the LMC and one in the SMC). These detections lead to a lower limit of the SB1 fraction of Cepheids of $\sim 3/12= 25\%$ in the LMC and $(5 + 2 \rm{(not\ analyzed)})/26 \sim 27\%$ for the SMC\footnote{Five detections among the 24 analyzed targets, plus SMC-CEP-1686 and -1693, excluded precisely because of their strong orbital signatures.}. However, we note that the Magellanic Clouds harbor a large population of double-lined spectroscopic binaries not seen in the Milky Way \citep{Pilecki2021,Pilecki2025}. Detailed comparisons between the binary fractions of the Magellanic Clouds and the Milky Way, where $29\%$ of Cepheids reside in SB1 systems (see \veloce-II), will further benefit from our methodology and the upcoming 4MOST observations. Additionally, the spatial distribution of $v_\gamma$ values determined for LMC Cepheids matches the expected line-of-sight component of its large-scale rotation. In the SMC, the situation appears to be more complex.

Our PCA-based framework provides a robust foundation for systematic investigations of Cepheid multiplicity and galaxy kinematics based on stellar sub-populations across the Milky Way and the Magellanic Clouds in the era of 4MOST and large ground-based spectroscopic surveys. Future work will also consider extending this approach to RV time-series measurements from {\it Gaia}. 

\begin{acknowledgements}
The authors acknowledge the referee's constructive comments, which have helped to improve this manuscript.
This work was supported by the European Research Council (ERC) under the European Union’s Horizon 2020 research and innovation programme (Grant Agreement No. 947660). RIA was funded by the SNSF through an Eccellenza Professorial Fellowship, grant number PCEFP2\_194638. Funded by zukunft.niedersachsen, the joint science funding program of the Lower Saxony Ministry of Science and Culture and the Volkswagen Foundation.
This research has made use of the VizieR catalogue access tool, CDS, Strasbourg Astronomical Observatory, France (DOI : 10.26093/cds/vizier) \citep{ochsenbein_vizier_2000}.
This research has made use of the SIMBAD database, CDS, Strasbourg Astronomical Observatory, France \citep{wenger_simbad_2000}. 
This research has made use of the Astrophysics Data System, funded by NASA under Cooperative Agreement 80NSSC21M0056.
This work has made use of data from the European Space Agency (ESA) mission {\it Gaia} (\href{https://www.cosmos.esa.int/gaia}{\nolinkurl{cosmos.esa.int/gaia}}), processed by the {\it Gaia} Data Processing and Analysis Consortium (DPAC, \href{https://www.cosmos.esa.int/web/gaia/dpac/consortium}{\nolinkurl{cosmos.esa.int/web/gaia/dpac/consortium}}). Funding for the DPAC has been provided by national institutions, in particular the institutions participating in the {\it Gaia} Multilateral Agreement.
The analysis was conducted using the \texttt{Python}\footnote{Python Software Foundation, \href{https://www.python.org/}{\nolinkurl{python.org}}} programming language and the related \texttt{Jupyter Notebook} application from \texttt{Project Jupyter}\footnote{Project Jupyter, \href{https://jupyter.org/}{\nolinkurl{jupyter.org}}}. Our work mainly used the following open-source libraries: \texttt{numpy} \citep{harris_array_2020}, \texttt{pandas} \citep{team_pandas-devpandas_2025}, \texttt{scikit-learn} \citep{pedregosa_scikit-learn_2011}, \texttt{statsmodels} \citep{seabold2010statsmodels}, \texttt{scipy} \citep{virtanen_scipy_2020},  \texttt{astropy} \citep{astropy_collaboration_astropy_2013, astropy_collaboration_astropy_2018, astropy_collaboration_astropy_2022}, \texttt{astroquery} \citep{ginsburg_astroquery_2019}, \texttt{emcee} \citep{foreman-mackey_emcee_2013}, \texttt{matplotlib} \citep{hunter_matplotlib_2007}, and \texttt{dill} \citep{mckerns_building_2012, pathos_2010}. We thank the developers for developing, maintaining, and openly distributing these libraries.
The authors thank the organizers of the RRLCEP-RR Lyrae \& Cepheid Stars Conference 2024 (hosted  by Cadi Ayyad University and Oukaimeden Observatory in Marrakesh, Morocco) for granting us the possibility to present this work's preliminary results.

\end{acknowledgements}

\bibliographystyle{bibtex/aa}
\bibliography{bibtex/biblio}

\begin{appendix}

\section{Examples of reconstructed RV curves}\label{appendix:examples_reconstructed}
\begin{figure*}
    \centering
    \includegraphics[width=.95\textwidth]{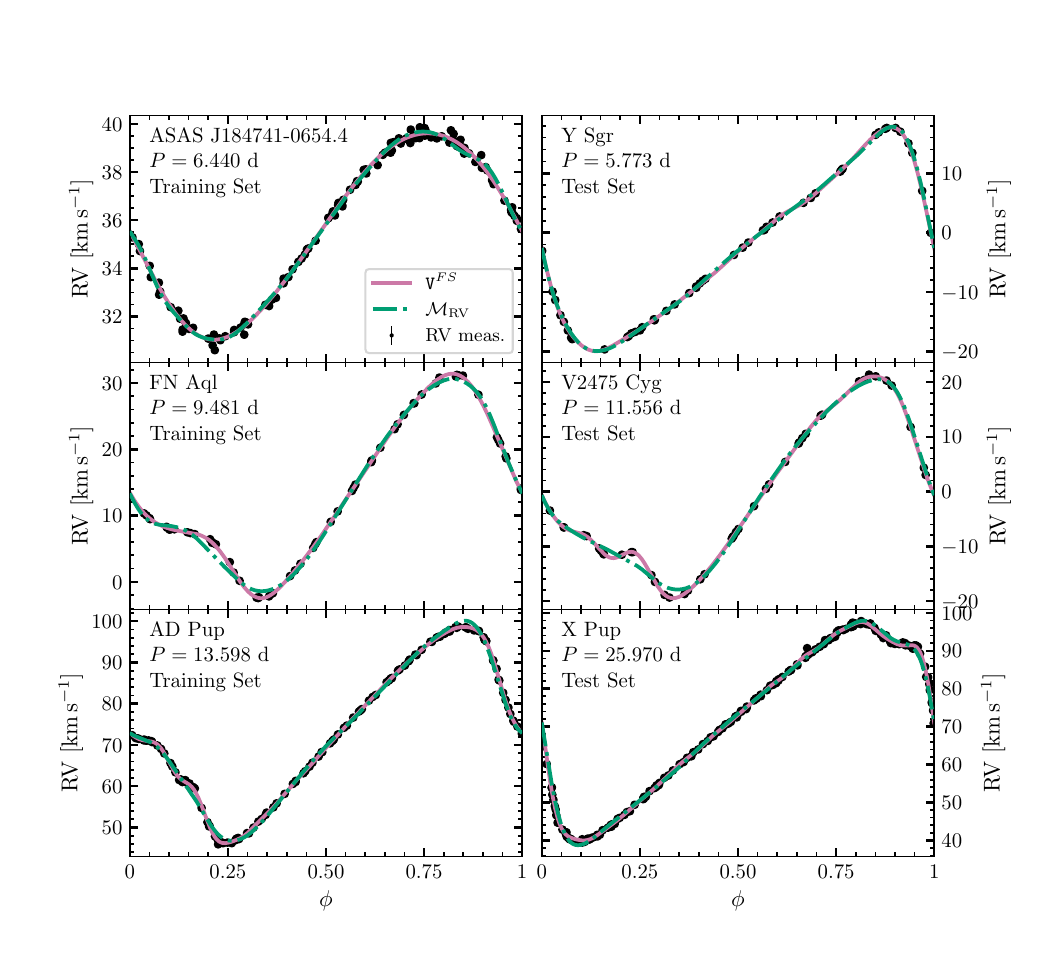}
    \caption{\label{fig:example_fit_rv} Examples of RV curves reconstructed by fitting \veloce\ RV measurements using PCs. The targets on the left panels are members of the training set, while those on the right panels belong to the test set. The pulsation period increases from top to bottom in each column. $\texttt{V}^{\rm FS}$ and $\mathcal{M}_{\rm RV}$ are shown in pink and green, respectively, whereas \veloce\ RV measurements are presented in black (uncertainties are shown, although they are often smaller than the symbol).}
\end{figure*}

Figure \ref{fig:example_fit_rv} shows RV curves reconstructed by fitting the complete set of \veloceDRI\ RV measurements, for 6 Cepheids spanning the period range of the sample: 3 belonging to the training set and 3 to the test set (cf. Section \ref{subsec:pca_evaluation}).

\section{Results of the application of the framework to varying number of RV measurements}

In Section\,\ref{subsec:result_framework}, we applied our framework (the PCs with the determined priors) to randomly selected $N_{\rm RV}$ measurements for each analyzed \veloceDRI\ Cepheid. The analysis was repeated for $N_{\rm RV}$ values ranging from 3 to 20, with results reported in Tab. \ref{tab:nrv_summary} and in Fig. \ref{fig:subsampling_3} (for $N_{\rm RV} = 3$). Here, we present the corresponding figures for $N_{\rm RV} = 4,\ 5,\ 7$ and 10, together with a few examples of successful and unsuccessful reconstructions. Figure\,\ref{fig:example_fit_nrv} presents a few examples of RV curves reconstructed by fitting $N_{\rm RV}=3$ measurements.

\begin{figure*}[ht!]
    \centering
    \includegraphics[width=0.48\textwidth]{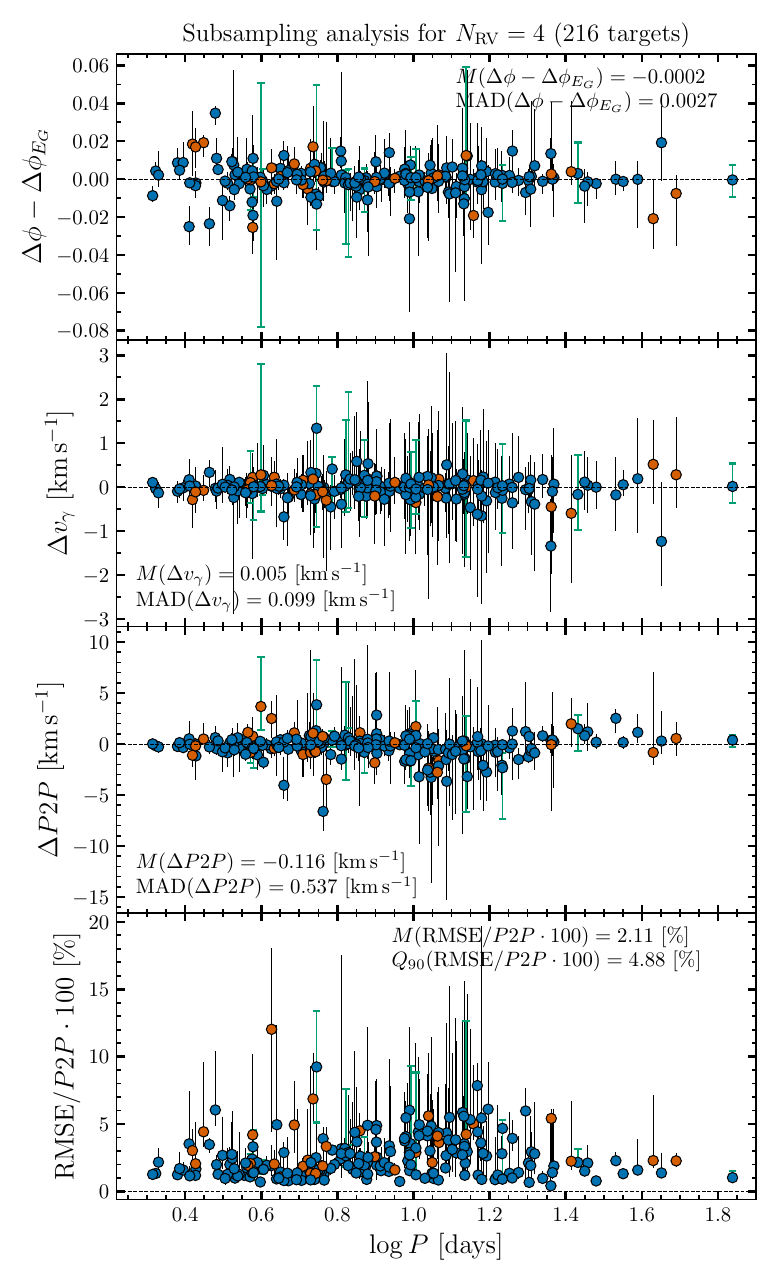}
    \includegraphics[width=0.48\textwidth]{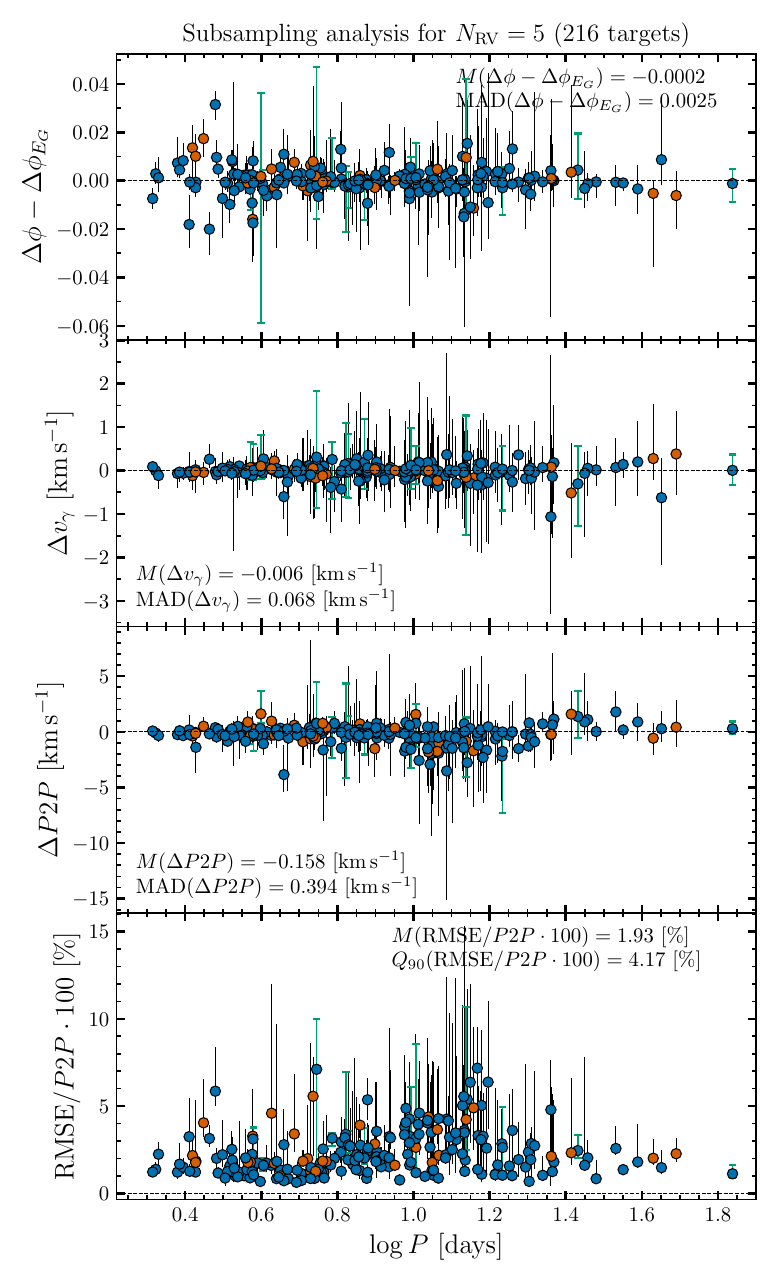}

    \caption{Performance of the PCs and priors applied to subsamples containing $N_{\rm RV} = 4$ and 5 RV measurements. See Figure\,\ref{fig:subsampling_3}. \label{fig:subsampling_examples_1}}
\end{figure*}

\begin{figure*}[ht!]
    \centering
    \includegraphics[width=0.48\textwidth]{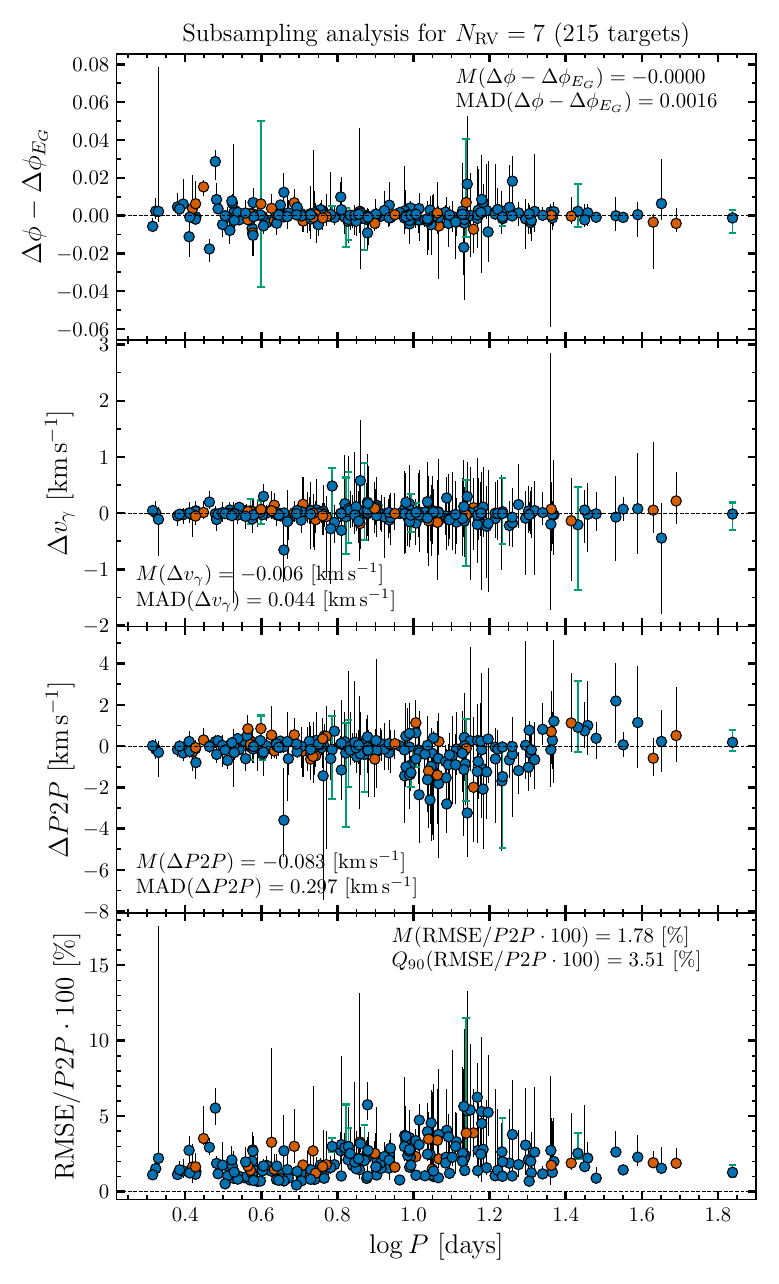}
    \includegraphics[width=0.48\textwidth]{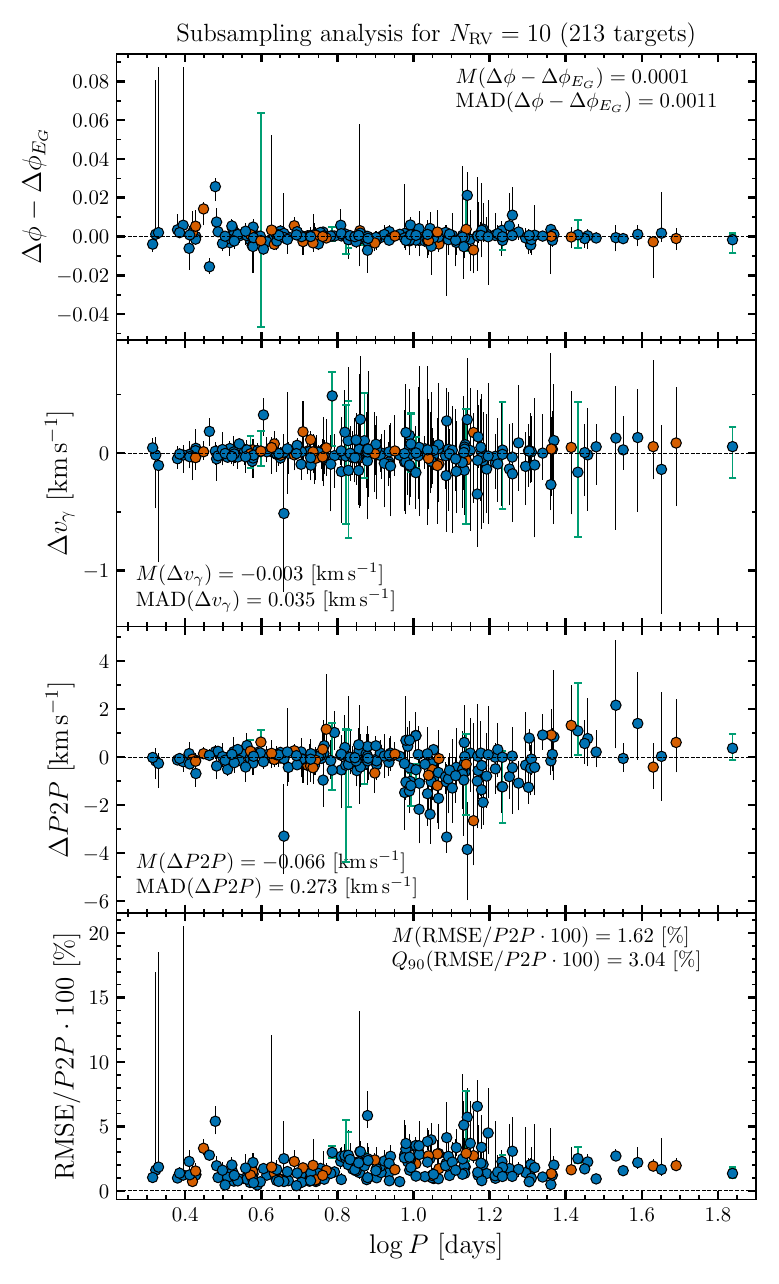}
    \caption{Performance of the PCs and priors applied to subsamples containing $N_{\rm RV} = 7$ and 10 RV measurements. See Figure\,\ref{fig:subsampling_3}. \label{fig:subsampling_examples_2}}
\end{figure*}

\begin{figure*}
    \centering
    \includegraphics[width=.95\textwidth]{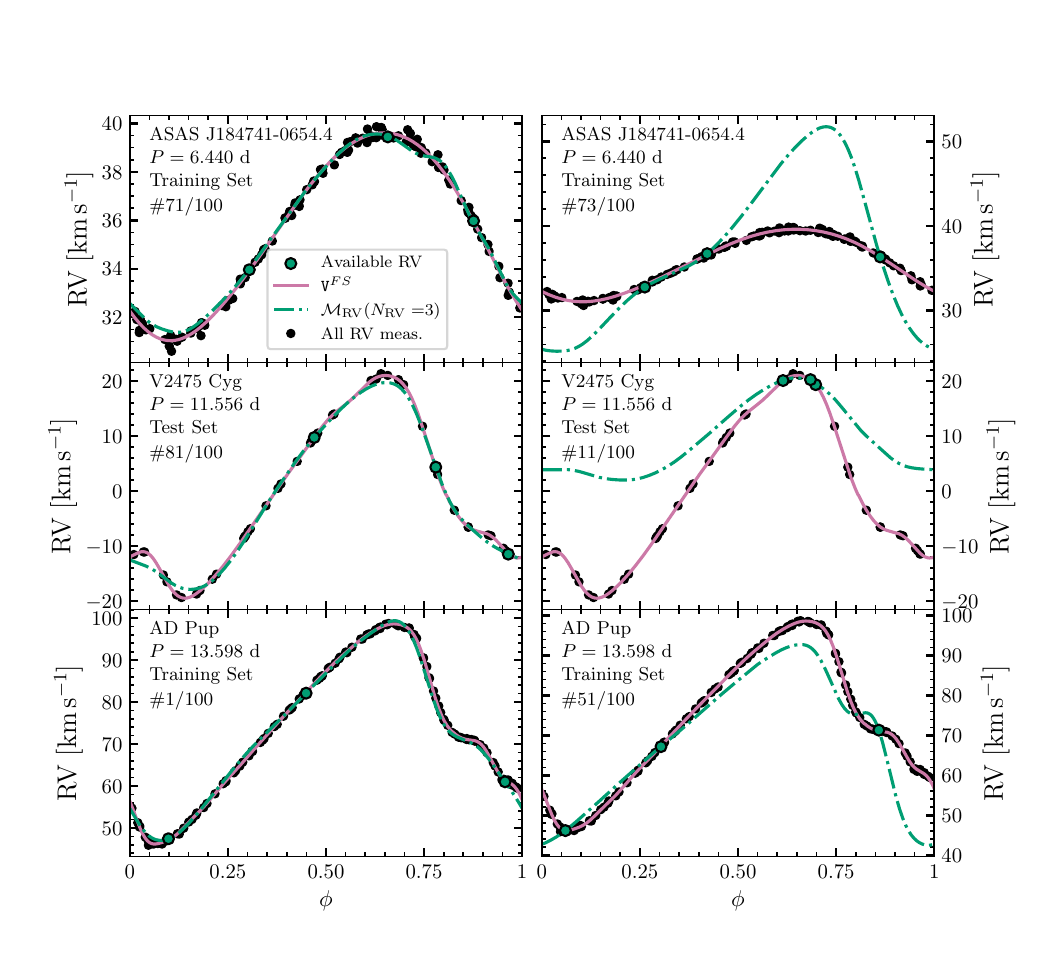}
    \caption{\label{fig:example_fit_nrv} Examples of RV curves reconstructed by fitting $N_{\rm RV} = 3$ RV measurements using our framework. Each row shows the same target for two separate subsamples: the left panels show an example of a successful reconstruction, the right panels one of the worst reconstructions obtained for that target. $\texttt{V}^{\rm FS}$ and $\mathcal{M}_{\rm RV}(N_{\rm RV} = 3)$ are shown in pink and green, respectively. All \veloce\ RV measurements are plotted in black with uncertainties shown, although often smaller than the symbols. The $N_{\rm RV} = 3$ measurements used for the reconstruction are highlighted in green. The target and subsample number are indicated in each panel.}
\end{figure*}

\section{Results for the Magellanic Clouds}
In Section\,\ref{subsec:extragalactic}, we applied our framework to RV measurements of Cepheids in the LMC and SMC, identifying SB1 candidates within the datasets. Here, we present the RV curve properties derived by fitting the $\mathcal{M}_{\rm RV}^{\rm SOLO}$ and  $\mathcal{M}_{\rm RV}^{\rm MULTI}$ models.
\begin{table}[]
    \small    \centering
    \caption{Extragalactic Cepheids RV curve properties \label{tab:extragalactic}}
    \begin{tabular}{@{}lc@{}c@{}cc@{}}

    \toprule
        \multicolumn{5}{c}{\citet{storm_calibrating_2011} LMC targets }\\
    \midrule
     Target & $\overline{v_{\gamma, \rm cluster}}$ & $P2P(v_{\gamma, \rm cluster})$ & $P2P$ & $RMSE$ \\
     &[\kms] &  [\kms]   &  [\kms]  &  [\kms] \\
    \midrule
  HV873           & $251.21 \pm 0.02$ & -             & $50.29$ & $1.17\ (2.3\%)$ \\
  HV876           & $296.47 \pm 0.02$ & -             & $62.31$ & $0.82\ (1.3\%)$ \\
  HV877           & $292.87 \pm 0.02$ & -             & $29.49$ & $0.30\ (1.0\%)$ \\
  HV878           & $253.06 \pm 0.02$ & -             & $59.78$ & $0.68\ (1.1\%)$ \\
  HV881           & $242.36 \pm 0.02$ & -             & $52.63$ & $0.88\ (1.7\%)$ \\
  HV900           & $235.24 \pm 0.02$ & -             & $47.19$ & $0.56\ (1.2\%)$ \\
 HV914$\dagger$  & $258.05 \pm 0.77$ & 3.91          & $40.33$ & $1.06\ (2.6\%)$ \\
 HV1005$\dagger$ & $273.67 \pm 2.67$ & 11.34         & $63.55$ & $0.83\ (1.3\%)$ \\
  HV1006          & $271.75 \pm 0.03$ & -             & $52.73$ & $0.78\ (1.5\%)$ \\
  HV1023          & $291.64 \pm 0.02$ & -             & $53.2 $ & $2.45\ (4.6\%)$ \\
  HV2282          & $256.79 \pm 0.03$ & -             & $54.57$ & $0.67\ (1.2\%)$ \\
  HV2369          & $308.11 \pm 0.03$ & -             & $50.3 $ & $0.69\ (1.4\%)$ \\
  HV2405          & $210.19 \pm 0.02$ & -             & $26.71$ & $0.97\ (3.6\%)$ \\
  HV2527          & $230.44 \pm 0.03$ & -             & $53.11$ & $0.77\ (1.4\%)$ \\
  HV2538          & $278.29 \pm 0.03$ & -             & $33.09$ & $0.73\ (2.2\%)$ \\
  HV2549          & $237.91 \pm 0.03$ & -             & $39.65$ & $1.27\ (3.2\%)$ \\
  HV5655          & $222.00 \pm 0.03$ & -             & $53.65$ & $0.61\ (1.1\%)$ \\
  HV6093          & $277.20 \pm 0.02$ & -             & $42.39$ & $0.52\ (1.2\%)$ \\
  HV12452         & $234.23 \pm 0.02$ & -             & $38.87$ & $0.63\ (1.6\%)$ \\
  HV12505         & $284.73 \pm 0.03$ & -             & $53.38$ & $1.03\ (1.9\%)$ \\
  HV12717         & $258.23 \pm 0.02$ & -             & $34.96$ & $0.81\ (2.3\%)$ \\
 U1$\dagger$     & $252.26 \pm 1.69$ & 6.29          & $61.44$ & $0.59\ (1.0\%)$ \\

\midrule
    \multicolumn{5}{c}{\citet{gieren_effect_2018} SMC targets }\\
    \midrule
 SMC-CEP-0320          & $164.98 \pm 0.01$ & -             & $13.43$ & $0.33\  (2.4\%)$ \\
 SMC-CEP-0518          & $124.65 \pm 0.02$ & -             & $52.73$ & $1.40\  (2.7\%)$ \\
 SMC-CEP-0524          & $145.25 \pm 0.03$ & -             & $32.28$ & $1.17\  (3.6\%)$ \\
 SMC-CEP-0958          & $112.32 \pm 0.01$ & -             & $55.66$ & $2.13\  (3.8\%)$ \\
 SMC-CEP-1365          & $148.14 \pm 0.01$ & -             & $22.39$ & $0.65\  (2.9\%)$ \\
 SMC-CEP-1385          & $123.70 \pm 0.02$ & -             & $36.03$ & $1.09\  (3.0\%)$ \\
 SMC-CEP-1403          & $149.95 \pm 0.01$ & -             & $32.27$ & $0.68\  (2.1\%)$ \\
 SMC-CEP-1410          & $150.14 \pm 0.03$ & -             & $47.48$ & $1.88\  (4.0\%)$ \\
 SMC-CEP-1680$\dagger$ & $110.51 \pm 1.41$ & 9.37          & $42.66$ & $1.44\  (3.4\%)$ \\
 SMC-CEP-1712          & $133.32 \pm 0.01$ & -             & $27.97$ & $0.75\  (2.7\%)$ \\
 SMC-CEP-1717          & $122.40 \pm 0.02$ & -             & $38.06$ & $1.16\  (3.0\%)$ \\
 SMC-CEP-1723          & $128.74 \pm 0.02$ & -             & $51.56$ & $0.73\  (1.4\%)$ \\
 SMC-CEP-1729$\dagger$ & $195.46 \pm 1.82$ & 11.36         & $51.17$ & $2.17\  (4.2\%)$ \\
 SMC-CEP-1750          & $162.80 \pm 0.01$ & -             & $57.29$ & $0.87\  (1.5\%)$ \\
 SMC-CEP-1761          & $142.36 \pm 0.02$ & -             & $55.46$ & $1.41\  (2.5\%)$ \\
 SMC-CEP-1765          & $166.73 \pm 0.01$ & -             & $29.81$ & $0.80\  (2.7\%)$ \\
 SMC-CEP-1797          & $152.36 \pm 0.01$ & -             & $31.67$ & $0.85\  (2.7\%)$ \\
 SMC-CEP-1977$\dagger$ & $157.14 \pm 2.21$ & 11.33         & $19.19$ & $1.32\  (6.9\%)$ \\
 SMC-CEP-2533          & $180.89 \pm 0.02$ & -             & $34.06$ & $1.66\  (4.9\%)$ \\
 SMC-CEP-2905$\dagger$ & $149.50 \pm 0.56$ & 4.24          & $30.99$ & $1.13\  (3.7\%)$ \\
 SMC-CEP-3311          & $180.12 \pm 0.01$ & -             & $32.92$ & $0.79\  (2.4\%)$ \\
 SMC-CEP-3927          & $186.67 \pm 0.02$ & -             & $51.18$ & $1.39\  (2.7\%)$ \\
 SMC-CEP-4017$\dagger$ & $185.25 \pm 0.68$ & 4.66          & $44.99$ & $1.90\  (4.2\%)$ \\
 SMC-CEP-4444          & $172.32 \pm 0.01$ & -             & $56.82$ & $1.04\  (1.8\%)$ \\

    \midrule
    \bottomrule
    \end{tabular}
    \tablefoot{Results of the application of the PC-based models to extragalactic Cepheids. The symbol $\dagger$ highlights targets exhibiting evidence of orbital motion according to our analysis; for these stars, we report the results of the $\mathcal{M}_{\rm RV}^{\rm MULTI}$ fit and the $P2P$ of the measured cluster-averaged velocities, $P2P(v_{\gamma, \rm cluster})$. For the rest of the targets, the results of the $\mathcal{M}_{\rm RV}^{\rm SOLO}$ fit are listed. The $RMSE$ column gives, in parentheses, the $RMSE$-to-$P2P$ ratio, indicating the relative strength of the residual scatter.}
\end{table}

\end{appendix}
\end{document}